\documentclass[twocolumn,aps,superscriptaddress]{revtex4-1}

\usepackage{amsfonts}
\usepackage{amsmath}
\usepackage{amssymb}
\usepackage[pdftex]{graphicx}
\usepackage{dcolumn}
\usepackage{bm}
\usepackage{braket}
\usepackage{color}
\usepackage{here}
\usepackage{mathrsfs}
\usepackage{mathtools}
\usepackage[normalem]{ulem}

\begin{document}

\title{Heat transport through a two-level system under continuous quantum measurement}

\author{Tsuyoshi Yamamoto}
\email{yamamoto.tsuyoshi.ga@u.tsukuba.ac.jp}
\author{Yasuhiro Tokura}
\affiliation{Faculty of Pure and Applied Sciences, University of Tsukuba, Tsukuba, Ibaraki 305-8571, Japan}
\author{Takeo Kato}
\affiliation{Institute for Solid State Physics, the University of Tokyo, Kashiwa, Chiba 277-8581, Japan}

\date{\today}

\begin{abstract}
We study the backaction of quantum measurements on heat transport through a two-level system by considering the continuous quantum measurement onto an eigen state of the two-level system.
For the nonselective measurement, the backaction appears as a dephasing effect on the two-level system.
We formulate the heat current under the selective measurement with a stochastic master equation and show that the cross-correlation between the measurement outcomes and the heat current contains information on the backaction.
We expect that our findings can be verified by using a platform of superconducting circuits.
\end{abstract}

\pacs{Valid PACS appear here}

\maketitle


\section{introduction}
Measurement is indispensable to access the information on a system.
Quantum measurements usually disturb a quantum system and thus destroy quantum correlations, in striking contrast to the classical realm in which unperturbed measurement is possible in principle~\cite{Wiseman_text}.
This backaction of the quantum measurement triggers many intriguing phenomena.
For example, in quantum many-body systems constituting large-scale entanglements, the quantum backaction induces various nontrivial effects, e.g., measurement-induced phase transitions~\cite{Li2018,Li2019,Skinner2019,Ippoliti2021,Minato2022}, non-Hermitian dynamics~\cite{Lee2014,Ashida2016,Ashida2018}, and suppression of the Kondo effect~\cite{Hasegawa2022}.
Moreover, quantum backaction has been observed in well-controllable systems of cold atoms~\cite{Murch2008,Syassen2014,Zhu2014,Patil2015,Tomita2017} and of solid-state nanostructures~\cite{Bischoff2015}.
In particular, a superconducting circuit is an ideal platform for exploring quantum backaction on many-body states.
This is because experimental efforts toward the realization of quantum computers have enabled high-speed readout of qubits' information and controlled interaction of a single qubit with other qubits or electromagnetic fields~\cite{Hatridge2013,Groen2013}.
{Recently, thermal engines under quantum measurement have been proposed theoretically in light of superconducting circuits~\cite{Campisi2017,Bhandari2022}.

Continuous measurement has a significant impact on transport through a small quantum object since transport properties reflect quantum states of such an object~\cite{Bernad2010, Rech2011}.
The simplest example is heat transport through a two-level system, i.e., a qubit.
Recent developments in experimental techniques have enabled us to accurately measure heat current through a qubit and have stimulated theoretical and experimental studies~\cite{Meschke2006,Timofeev2009,Partanen2016,Ronzani2018,Senior2020,Maillet2020,Pekola2021}.
Contrary to its apparent simplicity, quantum transport via a qubit exhibits many-body effects due to strong qubit-bath coupling, e.g., the Kondo effect~\cite{Leggett1987,Weiss_text,Hewson_text,Guinea1985,LeHur2012,Saito2013,Yamamoto2018NJP} and quantum phase transitions~\cite{Leggett1987,Weiss_text,Anderson1971,Kosterlitz1976,Filippis2022,Kehrein1995,Kehrein1996,Bulla2003,Winter2009,Yamamoto2018PRB}.
Moreover, since the transferred heat is related to entropy production~\cite{Pekola2015,Pekola2021}, heat transport under measurement is also a key to demonstrating Maxwell’s demon and shedding light on the relation between energy and information~\cite{Szilard1929,Leff_text,Sagawa2008,Sagawa2010,Koski2014,Koski2014PRL}.
In that sense, it is cross-cutting to consider measurement effects in heat transport from the viewpoint of uniting condensed matter physics and information theory.

In this paper, we consider heat transport through a two-level system and examine how heat transport is affected by a continuous quantum measurement onto an eigenstate of the two-level system (see Fig.~\ref{fig:system}).
We calculate the heat current by using the master equation which takes quantum measurement processes followed by postselection into account.
First, we show that the dephasing induced by continuous measurement modifies the heat current.
To clarify the backaction in detail, we further calculate the cross-correlation between the heat current and the measurement outcomes in the selective measurement scheme.
By introducing this quantity, it becomes possible to discuss the backaction to nonequilibrium transport phenomena more directly in feasible experimental setups.
We expect that the present experimental techniques enable the measurement setup to be realized in the form of superconducting circuits.

\begin{figure}
    \centering
    \includegraphics[width=0.8\columnwidth]{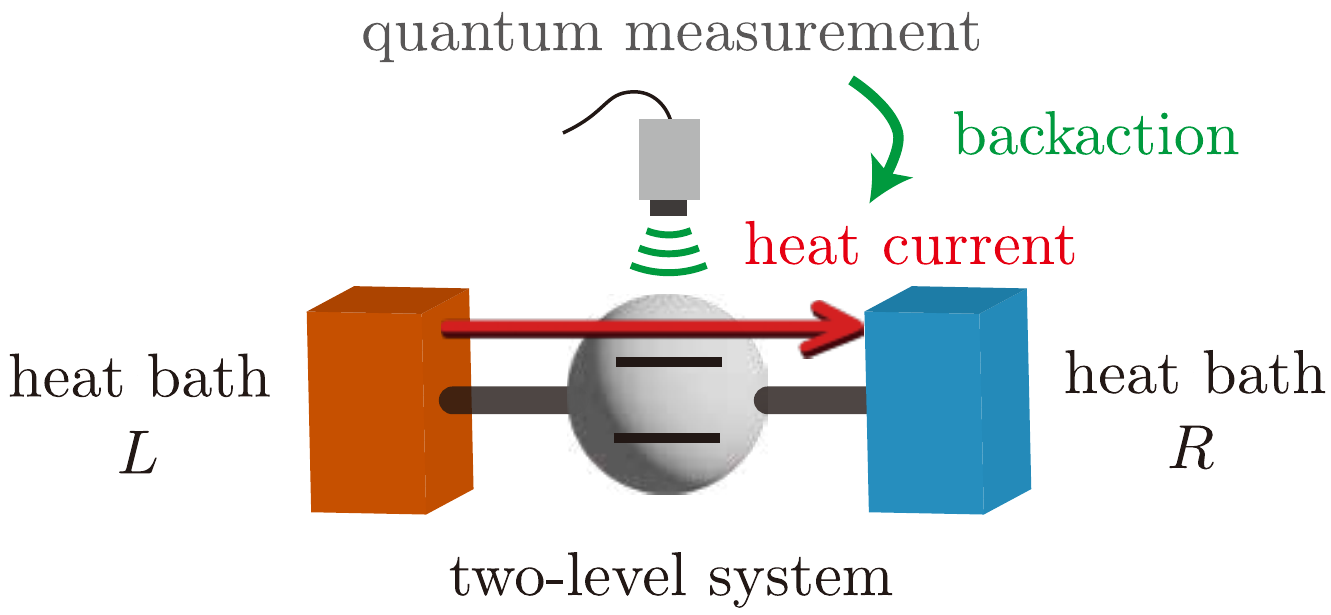}
    \caption{Schematic diagram of the model considered in this paper. A two-level system is coupled to two bosonic heat baths and an apparatus for continuous quantum measurement.
    When the temperature of the heat bath $L$ is higher than that of the heat bath $R$, the heat current flows from the heat bath $L$ to $R$ through the two-level system.
    The quantum measurement of the two-level system affects the heat current, which is called {\it backaction}.}
    \label{fig:system}
\end{figure}

\section{Model}

We consider heat current between two heat baths through a two-level system under continuous quantum measurement (see Fig.~\ref{fig:system}).
The system is described by the spin-boson model, whose Hamiltonian consists of three terms: $H=H_\mathrm{TLS}+H_\mathrm{B}+H_\mathrm{I}$.
The two-level system is described by 
\begin{align}
H_\mathrm{TLS}=-\frac{\hbar\Delta}{2}\sigma_x,
\end{align}
where $\Delta$ ($>0$) is the tunneling amplitude and $\sigma_i$ ($i=x,y,z$) are the Pauli matrices.
The heat baths are modeled by collections of harmonic oscillators as
\begin{align}
H_\mathrm{B}=\sum_{r}H_{\mathrm{B},r}=\sum_{rk}\hbar\omega_{rk}b_{rk}b_{rk}^\dagger ,
\end{align}
where $b_{rk}$ ($b_{rk}^\dagger$) is a bosonic annihilation (creation) operator of the $k$th mode in the heat bath $r$ ($=L,R$) with energy $\hbar \omega_{rk}$.
The interaction between the two-level system and the heat baths is described by
\begin{align}
H_\mathrm{I}=\sum_{r}H_{\mathrm{I},r}=-\frac{\sigma_z}{2}\sum_{rk}\hbar\lambda_{rk}\left(b_{rk}+b_{rk}^\dagger\right),
\end{align}
with coupling strength $\lambda_{rk}$.
The properties of the heat baths are determined by the spectral density $I_r(\omega)\equiv\sum_{k}\lambda_{rk}^2\delta(\omega-\omega_{rk})$.
In this paper, we focus on an Ohmic heat bath whose spectral density is written in the form~\cite{Leggett1987,Weiss_text}
\begin{align}
I_r(\omega)=2\alpha_r\omega e^{-\omega/\omega_\mathrm{c}},
\end{align}
where $\alpha_r$ is a dimensionless coupling strength and $\omega_\mathrm{c}$ is a cutoff frequency.

We consider the backaction induced by a quantum measurement onto the ground state $\ket{+}$ (or onto the excited state $\ket{-}$) of the two-level system, where $\ket{\pm}$ are eigenstates of $\sigma_x$ ($\sigma_x\ket{\pm}=\pm\ket{\pm}$).
We define projection operators, $P_{\pm}=\ket{\pm}\bra{\pm}=(I\pm\sigma_x)/2$, and operators for the corresponding continuous quantum measurements,
\begin{align}
M_{\pm}=\sqrt{\gamma_m^{\pm}}~P_{\pm},
\end{align}
where $\gamma_m^{\pm}$ indicates the strength of the measurement.

\section{Nonselective measurement}

\subsection{Quantum master equation}

First, let us consider a nonselective quantum measurement in which the measured values do not affect the subsequent dynamics.
In this case, the quantum dynamics can be described by the Lindblad equation after taking the ensemble average over the measurement outcomes~\cite{Carmichael_text}:
\begin{align}
\frac{d\rho}{dt}=-\frac{i}{\hbar}\left[H,\rho\right]+\mathcal{D}_m[\rho],
\end{align}
where $\mathcal{D}_m[\rho]=\sum_{i=\pm}(M_i\rho M_i^\dagger-\{M_i^\dagger M_i,\rho\}/2)$.
Here, $\rho$ denotes the density matrix and $[\cdot,\cdot]$ and $\{\cdot,\cdot\}$ are the commutator and anticommutator, respectively.
Assuming that the system-bath coupling is weak ($\alpha_r\ll1$), the Lindblad equation for the reduced density matrix, $\tilde{\rho}=\mathrm{tr}_\mathrm{B}[\rho]$, is obtained as~\cite{Breuer_text}
\begin{align}
\label{eq:lindblad}
\frac{d\tilde{\rho}}{dt}
=-\frac{i}{\hbar}\left[H_\mathrm{TLS},\tilde{\rho}\right]+\mathcal{D}_\mathrm{B}[\tilde{\rho}]+\mathcal{D}_m[\tilde{\rho}],
\end{align}
where $\mathcal{D}_\mathrm{B}[\tilde{\rho}]=\sum_{r,i=\pm}\Gamma_{ri}(\sigma^i\tilde{\rho}\sigma^{-i}-\{\sigma^{-i}\sigma^i,\tilde{\rho}\}/2)$, $\sigma^\pm=(\sigma_z\pm i\sigma_y)/2$, $\Gamma_{r+}=(\pi/2)n_r(\Delta)I_r(\Delta)$ and $\Gamma_{r-}=(\pi/2)[n_r(\Delta)+1]I_r(\Delta)$ are the absorption and emission rates, respectively, and $n_r(\omega)=(e^{\beta_r\hbar\omega}-1)^{-1}$ is the Bose-Einstein distribution function at temperature $T_r=1/(k_\mathrm{B}\beta_r)$ of the heat bath $r$.
In the Lindblad equation~\eqref{eq:lindblad}, the effect of the measurement is described by $\mathcal{D}_m[\tilde{\rho}]=\gamma_m(\sigma_x\tilde{\rho}\sigma_x-\tilde{\rho})/4$ with $\gamma_m=\gamma_m^++\gamma_m^-$.
This indicates that the backaction of the nonselective measurement appears through dephasing~\cite{Gambetta2006}, whose amplitude is given by $\gamma_\mathrm{p}=\gamma_m/2$~\cite{Iyoda2013}.
Note that we cannot get the information on which state is detected in this scheme.

\subsection{Heat current}

The heat current flowing from the heat bath $L$ to $R$ through the two-level system is defined as $J_L=-dH_{\mathrm{B},L}/dt$.
Assuming that the temperature difference between the two heat baths is sufficiently small, i.e., $T_{L/R}=T\pm\delta T/2$ with $\delta T/T\ll1$, we can obtain an analytical formula for the steady-state heat current $\braket{J}=\braket{J_L}=-\braket{J_R}$ up to first order in $\delta T$ by using the Keldysh formalism, as~\cite{Ojanen2008,Saito2008,Saito2013}
\begin{align}
\label{eq:current_non-selective}
\braket{J}=\frac{\alpha\eta k_\mathrm{B}\delta T}{16}\int_0^\infty d\omega~S(\omega)\tilde{I}(\omega)\frac{(\beta\hbar\omega)^2}{\sinh(\beta\hbar\omega)},
\end{align}
where $\alpha=\alpha_L+\alpha_R$, $\eta=4\alpha_L\alpha_R/\alpha^2$, $\tilde{I}(\omega)=I_r(\omega)/\alpha_r$, and $\beta=1/(k_\mathrm{B}T)$.
Here, $S(\omega)$ is the Fourier transformation of a symmetrized correlation function, $S(t)=\langle \{\sigma_z(t),\sigma_z(0)\}\rangle/2$.
From the Lindblad equation~\eqref{eq:lindblad}, $S(\omega)$ reads \footnote{Using the quantum regression theorem~\cite{Breuer_text}, the dynamics of the correlation function  $C_{zz}(t)=\braket{\sigma_z(t)\sigma_z(0)}$ are determined by the differential equation, $\ddot{C}_{zz}+2\tilde{\Gamma}_\mathrm{s}\dot{C}_{zz}+(\Delta^2+\tilde{\Gamma}_\mathrm{s}^2)C_{zz}=0$, and its solution is given by $C_{zz}(t)=e^{-\tilde{\Gamma}_\mathrm{s}t}[\cos(\Delta t)+(\tilde{\Gamma}_\mathrm{s}/\Delta)\sin(\Delta t)]$ under the initial conditions $C_{zz}(0)=1$ and $\dot{C}_{zz}(0)=0$}
\begin{align}
\label{eq:Somega}
S(\omega)
=\frac{4\tilde{\Gamma}_\mathrm{s}(\Delta^2+\tilde{\Gamma}^2_\mathrm{s})}{[(\omega-\Delta)^2+\tilde{\Gamma}_\mathrm{s}^2][(\omega+\Delta)^2+\tilde{\Gamma}_\mathrm{s}^2]},
\end{align}
where $\tilde{\Gamma}_\mathrm{s}=(\Gamma_\mathrm{s}+\gamma_m)/2$ and $\Gamma_\mathrm{s}=\sum_r(\Gamma_{r+}+\Gamma_{r-})$
\footnote{The heat transport process discussed in this paper is called a sequential tunneling process, which can describe heat transport for $k_\mathrm{B}T\sim\hbar\Delta$. However, at lower temperatures, $k_\mathrm{B}T\lesssim0.1\hbar\Delta$, the sequential tunneling process is suppressed, and thus the co-tunneling process becomes dominant, in which the higher-order interaction between the two-level system and the heat baths plays the important role~\cite{Ruokola2011,Yamamoto2018NJP}.}.

\begin{figure}
    \centering
    \includegraphics[width=0.8\columnwidth]{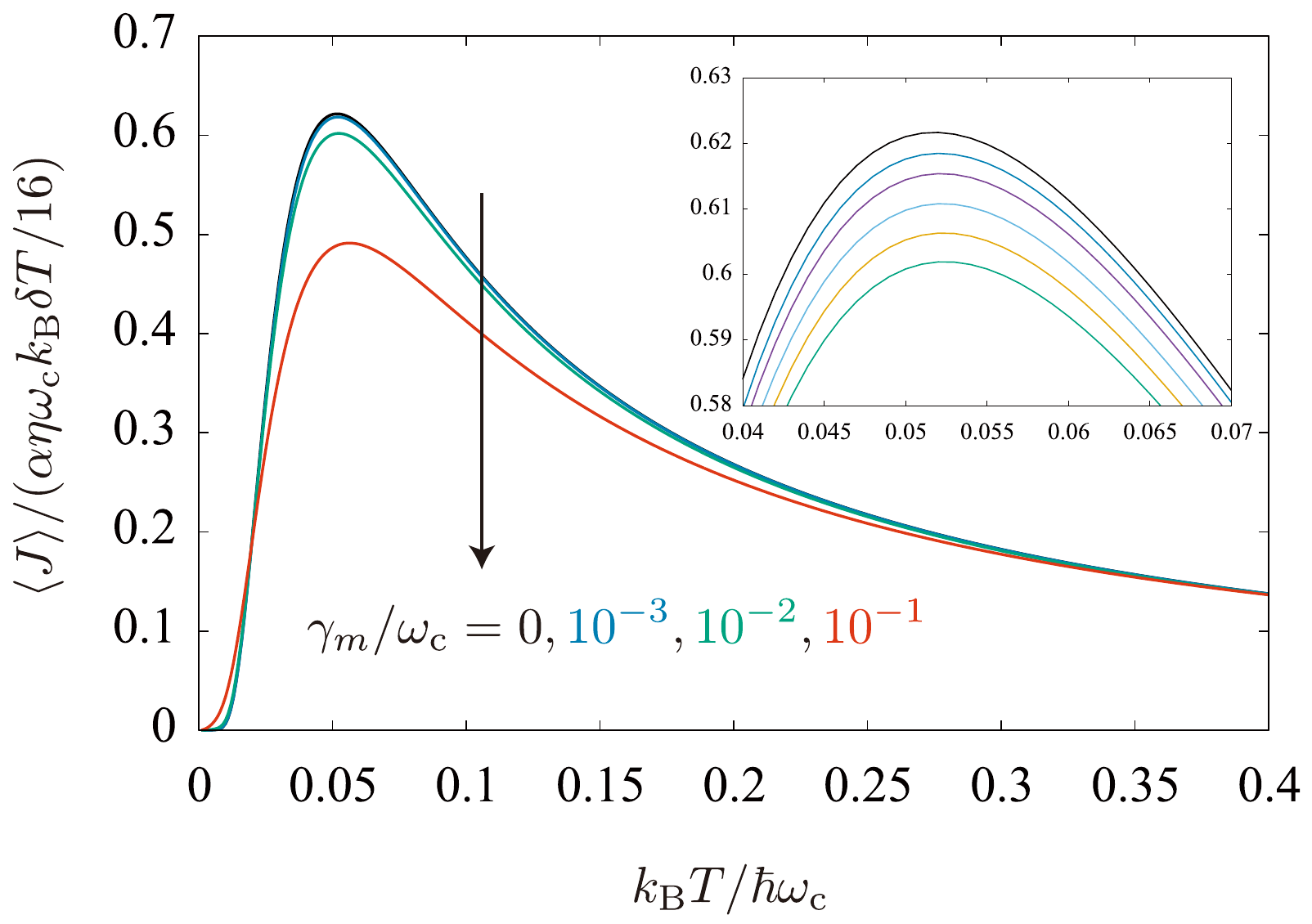}
    \caption{Temperature dependence of the steady-state heat current under a non-selective quantum measurement for $\Delta/\omega_\mathrm{c}=0.1$, $\alpha=0.01$, and $\gamma_m/\omega_\mathrm{c}=0$, $10^{-3}$, $10^{-2}$, and $10^{-1}$. The inset is an enlarged view near the peaks calculated for $\gamma_m/\omega_\mathrm{c}\times10^3=0$, 1, 2.5, 5, 7.5, and 10 from top to bottom.}
    \label{fig:current_non-selective}
\end{figure}

The steady-state heat current $\braket{J}$ is plotted as a function of temperature in Fig.~\ref{fig:current_non-selective}.
It shows a nonmonotonic behavior as a function of temperature.
This behavior can be described by {\it sequential tunneling process} in which heat is transported by a combination of energy absorption and emission accompanied by transitions between the ground and excited states of the two-level system~\cite{Segal2005}.
In this process, the heat current is  enhanced when the temperature is compatible with the energy splitting $\hbar\Delta$ of the two-level system.
As the strength of continuous quantum measurement, $\gamma_m$, increases, the peak of the heat current is suppressed.
As indicated, this suppression is induced by an increase in the level broadening in the response function~\eqref{eq:Somega} ($\Gamma_\mathrm{s}\to2\tilde{\Gamma}_\mathrm{s}$) due to the additional damping induced by the nonselective quantum measurement.

Here, we briefly explain how the nonselective measurement is related to the selective measurement discussed in Sec.~\ref{sec:selective}.
In general, the backaction of quantum measurement depends on its outcomes.
In the nonselective measurement, however, the backaction is averaged out for all the possible outcomes with their probabilities.
Figure~\ref{fig:current_selective} illustrates the relation between the heat current for the nonselective measurement and an original single-shot sample.
Before ensemble averaging, the heat current changes instantaneously according to the measurement outcomes at random times.
The set of times at which the heat current jumps is different from other ensembles, and the ensemble-averaged heat current is reduced to the one on the nonselective measurement.
Therefore, to access information on the backaction effect depending on the outcomes, we need to consider a physical quantity that represents the correlation between the outcome and the corresponding resultant heat current.
For this purpose, we will introduce the correlation function in the next section.

\section{Selective measurement}
\label{sec:selective}

\subsection{Stochastic quantum master equation}

Next, let us discuss selective measurement onto $\ket{i=\pm}$ by considering physical quantities which depend on measurement outcomes.
For this purpose, we treat discontinuous changes in the density matrix, i.e., quantum jumps using the stochastic master equation for a conditional density matrix $\rho_\mathrm{c}$~\cite{Wiseman1993,Wiseman_text}:
\begin{align}
\rho'_\mathrm{c}
=\rho_\mathrm{c}-\frac{i}{\hbar}\left[H,\rho_\mathrm{c}\right]\Delta t+\mathcal{D}_m^{(1)}\left[\rho_\mathrm{c}\right]\Delta t 
+\mathcal{D}_m^{(2)}\left[\rho_\mathrm{c}\right]\Delta N_i ,
\label{rhocdash}
\end{align}
where $\rho'_\mathrm{c}=\rho_\mathrm{c}(t+\Delta t)$, $\mathcal{D}_m^{(1)}[\rho_\mathrm{c}]=-\{M_i^\dagger M_i$, $\rho_\mathrm{c}\}/2+\mathrm{tr}[\rho_\mathrm{c}M_i^\dagger M_i]\rho_\mathrm{c}$,  $\mathcal{D}_m^{(2)}[\rho_\mathrm{c}]=M_i\rho_\mathrm{c}M_i^\dagger/\mathrm{tr}[\rho_\mathrm{c}M_i^\dagger M_i]-\rho_\mathrm{c}$, and we have neglected the higher-order terms of $\Delta t\Delta N\sim\mathit{o}(\Delta t)$.
The total density matrix is reproduced by taking the ensemble average, $\rho=E[\rho_\mathrm{c}]$, where $E[\cdot]$ denotes the ensemble average.
The measurement outcomes are described by a Poisson process with a stochastic variable $\Delta N_i$, which takes 1 when the two-level system is detected as being in state $\ket{i}$ and 0 otherwise during time $\Delta t$.
Note that $\Delta N_i=0$ does not mean that the quantum measurement is not performed: This means that the two-level system is measured not to be eigenstate $\ket{i}$ and the time evolution of the system is affected by $\mathcal{D}_m^{(1)}[\rho_\mathrm{c}]$ in Eq.~(\ref{rhocdash}).
The ensemble average of the Poisson variable is written as
\begin{align}
E[\Delta N_\pm]
=\mathrm{tr}[\rho_\mathrm{c}M_\pm^\dagger M_\pm]\Delta t
=\gamma_m^\pm\frac{1\pm\braket{\sigma_x}_\mathrm{c}}{2}\Delta t.
\end{align}

For the weak system-bath coupling ($\alpha_r\ll1$), the time-evolution equation for the reduced density matrix $\tilde{\rho}_\mathrm{c}=\mathrm{tr}_\mathrm{B}[\rho_\mathrm{c}]$ is obtained in the same manner as in the nonselective case, as 
\begin{align}
\tilde{\rho}'_\mathrm{c}
&=\tilde{\rho}_\mathrm{c}-\frac{i}{\hbar}\left[H_\mathrm{TLS},\tilde{\rho}_\mathrm{c}\right]\Delta t+\mathcal{D}_\mathrm{B}[\tilde{\rho}_\mathrm{c}]\Delta t \nonumber\\ &\qquad+\mathcal{D}_m^{(1)}[\tilde{\rho}_\mathrm{c}]\Delta t+\mathcal{D}_m^{(2)}[\tilde{\rho}_\mathrm{c}]\Delta N_i.
\end{align}
From this equation, one can derive time-evolution equations for the conditional coherences and population $\braket{\sigma_i(t)}_\mathrm{c}$ (see details in Appendix~\ref{app:sigma}).

\subsection{Heat current}

The heat current under the condition that the state $\ket{i}$ of the two-level system is not detected ($\Delta N_i=0$) is formulated as follows \footnote{In our simulation, the heat current under the condition, $\Delta N_i=1$, can be defined arbitrarily, because the number of time steps in this condition is much less than the number in $\Delta N_i=0$ in the limit $\Delta t\to0$. See Appendix~\ref{app:current-DN=0} for details.}.
In the Markov limit and assuming that the coherence can be neglected on the relevant time scale, the heat current flowing from the heat bath $r$ ($=L,R$) into the two-level system is expressed as~\cite{Segal2005} $\braket{J_r(t)}_\mathrm{c}=\hbar\Delta[\Gamma_{r+}p_\mathrm{c}^g(t)-\Gamma_{r-}p_\mathrm{c}^e(t)]$, where $p_\mathrm{c}^{g(e)}=\bra{g(e)}\tilde{\rho}_\mathrm{c}\ket{g(e)}=(1\pm\braket{\sigma_x}_\mathrm{c})/2$ is the population of the ground (excited) state.
For simplicity, we will consider a symmetric heat bath ($\alpha/2=\alpha_L=\alpha_R$) hereinafter.
When the temperature difference $\delta T$ is sufficiently small, the net heat current flowing from the heat bath $L$ to $R$, $\braket{J(t)}=(\braket{J_L(t)}-\braket{J_R(t)})/2$, can be expressed to first order in $\delta T$ as 
\begin{align}
\label{eq:current_selective}
\Braket{J(t)}_\mathrm{c}=\frac{\pi I(\Delta)k_\mathrm{B}\delta T}{8\,\mathrm{sinhc}^2(\beta\hbar\Delta/2)}\Braket{\sigma_x(t)}_\mathrm{c},
\end{align}
where $I(\Delta)=\sum_{r}I_r(\Delta)$ and $\mathrm{sinhc}(x)=\sinh(x)/x$.

\begin{figure}
    \centering
    \includegraphics[width=0.8\columnwidth]{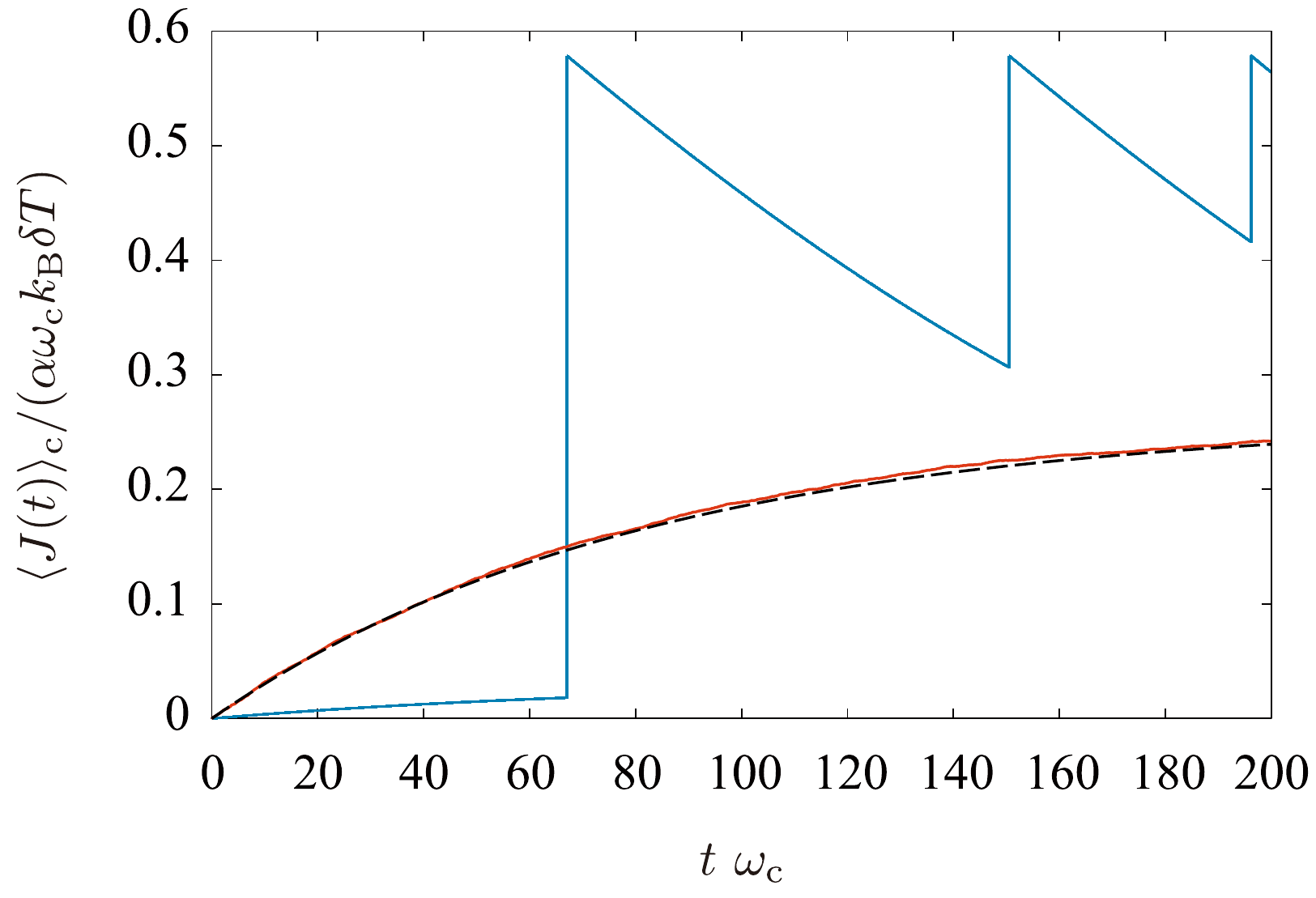}
    \caption{Heat current under quantum measurement onto $\ket{+}$ as a function of time. The parameters are $\Delta/\omega_\mathrm{c}=0.1$, $\alpha=0.01$, $\gamma_m^+/\omega_\mathrm{c}=0.01$, and $k_\mathrm{B}T/\hbar\omega_\mathrm{c}=0.1$.
    The blue line represents the heat current obtained from one sample of quantum trajectories which includes three quantum jumps in the range of the plot.
    The numerical ensemble average of the heat current, $E[\braket{J(t)}_\mathrm{c}]$, using $10^4$ quantum trajectories (the red line) coincides with the heat current obtained under the non-selective measurement (the black dashed line). We set the initial conditions as $\braket{\sigma_x(0)}_\mathrm{c}=\braket{\sigma_y(0)}_\mathrm{c}=0$ and $\braket{\sigma_z(0)}_\mathrm{c}=1$.}
    \label{fig:current_selective}
\end{figure}

A sample of the conditional heat current $\braket{J(t)}_\mathrm{c}$ under a continuous quantum measurement onto $\ket{+}$ is shown in Fig.~\ref{fig:current_selective}.
Several quantum jumps in the time evolution of $\braket{J(t)}_\mathrm{c}$ are characteristic to the selective measurement.
After detection, the conditional heat current jumps to a definite value because the state of the two-level system is projected to the state $\ket{+}$.
In contrast, the heat current tends to relax to a stationary point during periods in which there were no detections ($\Delta N_+=0$).
Thus the dynamics of the conditional heat current can be described by stochastic quantum trajectories composed of continuous dynamics during the no-detection periods ($\Delta N_+=0$) and quantum jumps ($\Delta N_+=1$). 
The conditional heat current under selective measurement onto $\ket{-}$ can be formulated in a similar way.
The ensemble average of the conditional heat current on these trajectories, $E[\braket{J(t)}_\mathrm{c}]$, (the red line in Fig.~\ref{fig:current_selective}) reproduces the heat current under the nonselective measurement (the dashed line in Fig.~\ref{fig:current_selective}) \footnote{We can reproduce this expression for the heat current at $t\to\infty$ by plugging $S(\omega)\approx\pi\delta(\omega-\Delta)$ into Eq.~\eqref{eq:current_non-selective}},
\begin{align}
\braket{J(t)}
=\pi I(\Delta) k_\mathrm{B}\delta T\frac{(\beta\hbar\Delta)^2}{16\sinh(\beta\hbar\Delta)}\left(1-e^{-\Gamma_\mathrm{s}t}\right).
\end{align}

\subsection{Cross-correlation}

So far, we focused on the steady-state heat current.
From now on, let us consider to what extent the measurement outcomes and the heat current are correlated in time to discuss the backaction due to the continuous quantum measurement onto $\ket{i=\pm}$.
For this purpose, we introduce the cross-correlation between the measurement outcomes and the heat current defined by
\begin{align}
F_i(t)=E\left[\frac{1}{t_1-t_0}\int_{t_0}^{t_1}dt'~\delta\Delta N_i(t')\delta\braket{J(t+t')}_\mathrm{c}\right],
\end{align}
where $\delta A=A-E[A]$.
Using Eq.~\eqref{eq:current_selective}, the cross-correlation reads
\begin{align}
F_i(t)=\frac{\pi I(\Delta)k_\mathrm{B}\delta T}{8\mathrm{sinhc}^2(\beta\hbar\Delta/2)}\mathcal{F}_i(t),
\end{align}
where
\begin{align}
\mathcal{F}_i(t)
&=\frac{1}{t_1-t_0}\int_{t_0}^{t_1}dt'~\Big\{E[\Delta N_i(t')\braket{\sigma_x(t+t')}_\mathrm{c}] \nonumber\\
&\qquad\qquad\qquad-E[\Delta N_i(t')]E[\braket{\sigma_x(t+t')}_\mathrm{c}]\Big\}.
\label{calFit}
\end{align}
At large $t_0,~t_1\gg\Gamma_\mathrm{s}^{-1}$, we can replace $E[\braket{\sigma_x(t’)}_\mathrm{c}]$ with the stationary solution, $\braket{\sigma_x}_\mathrm{ss}=\tanh(\beta \hbar\Delta/2)$~\footnote{The stationary solution of $\braket{\sigma_x}$ is calculated by solving the differential equation $\braket{\dot{\sigma}_x}=-\Gamma_\mathrm{s}\braket{\sigma_x}-\Gamma_\mathrm{a}$ under the steady-state condition $\braket{\dot{\sigma}_x}=0$.}.
Moreover, the averaged interval time between adjacent measurement events ($\Delta N_i=1$) can be assumed to be much longer than the inverse of the decay rate of $\braket{\sigma_x(t)}$, i.e., $(E[\Delta N_i]/\Delta t)^{-1}\gg\Gamma_\mathrm{s}^{-1}$.
This assumption is justified in our setup, in which a detection is a rare event according to the Poisson process.
In this situation, the contributions from two different measurement events can be neglected after taking the ensemble average, and the first term in the integrand in Eq.~(\ref{calFit}) can be replaced with a product of the probability of $\Delta N_i(t')=1$ and $\braket{\sigma_x(t+t')}$ with the initial condition $\braket{\sigma_x(t')}=\pm 1$.
Finally, we obtain the analytic expression for the cross-correlation as
\begin{align}
\label{eq:cross_analytics}
\frac{F_\pm(t)}{\Delta t}
=\pm\gamma_m^{\pm}\frac{\pi I(\Delta)k_\mathrm{B}\delta T}{16\mathrm{sinhc}^2(\beta\hbar\Delta)}e^{-\Gamma_\mathrm{s}t}.
\end{align}
This expression indicates that the cross-correlations under the continuous quantum measurement onto $\ket{\pm}$ decay exponentially with the decay rate $\Gamma_\mathrm{s}$ and have opposite signs, i.e., $F_+(0^+)/\gamma_m^+=-F_-(0^+)/\gamma_m^->0$.

\begin{figure}
    \centering
    \includegraphics[width=0.8\columnwidth]{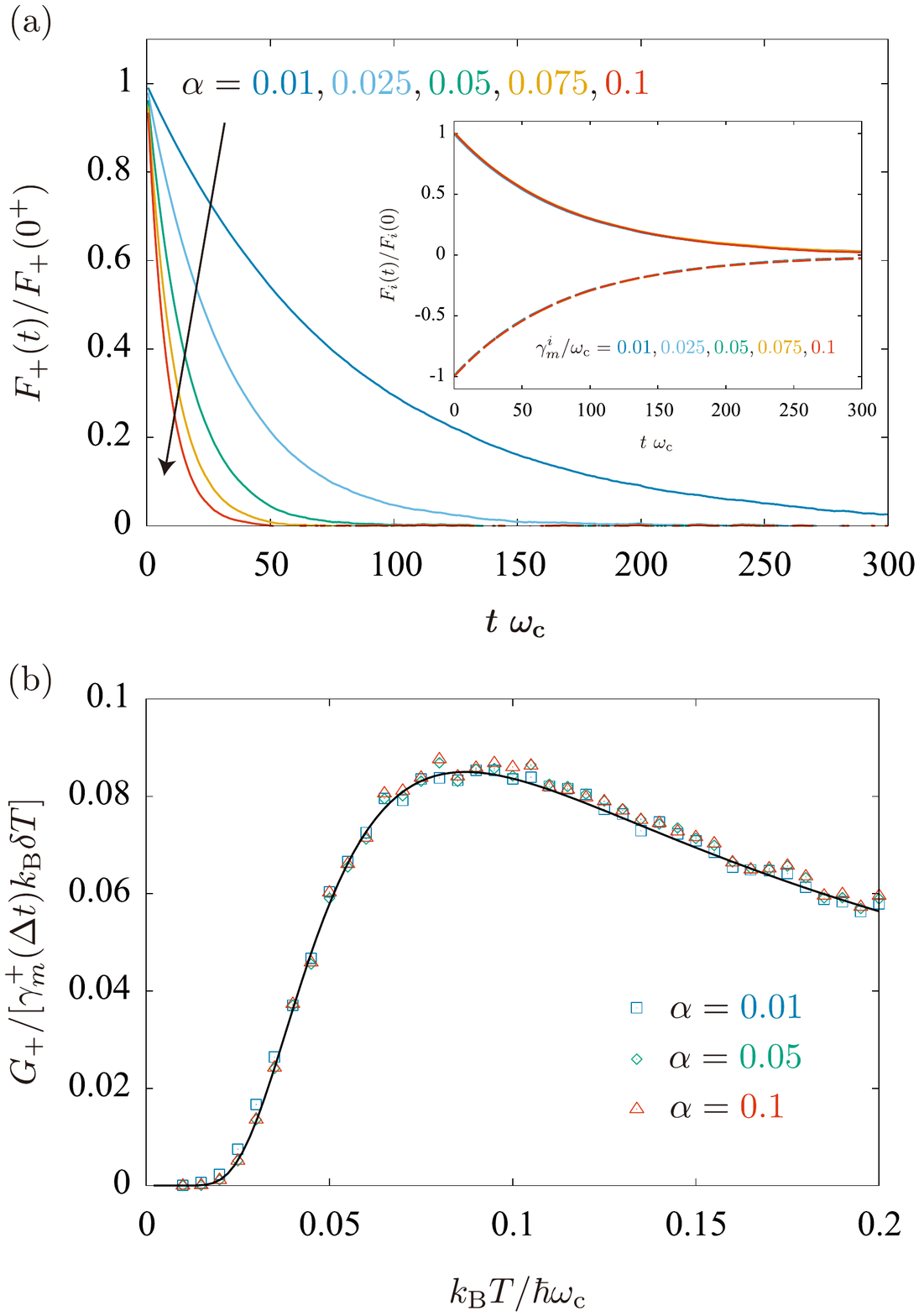}
    \caption{(a) Cross-correlation function $F_+(t)$ under selective measurement onto $\ket{+}$ for different values of $\alpha=0.01$, 0.025, 0.05, 0.075, and 0.1 using $10^4$ ensembles. The parameters are $\Delta/\omega_\mathrm{c}=0.1$, $\gamma_m^+=0.01$, $k_\mathrm{B}T/\hbar\omega_\mathrm{c}=0.1$, $t_0\omega_\mathrm{c}=500$, and $t_1\omega_\mathrm{c}=1000$. The inset shows the cross-correlation functions, $F_+(t)$ (solid lines) and $F_-(t)$ (dashed lines), for different $\gamma_m^i=0.01$, 0.025, 0.05, 0.075, and 0.1. 
    (b) Integrated cross-correlation $G_+$ as a function of temperature for $\alpha=0.01$, 0.05, and 0.1 using $5 \times 10^3$ ensembles. The other parameters are the same as in (a). The solid line represents an analytical expression~\eqref{eq:G_analytics}. For both panels, the initial conditions are the same as those in Fig.~\ref{fig:current_selective}.}
    \label{fig:cross-correlation}
\end{figure}

Figure~\ref{fig:cross-correlation}(a) shows the results of numerical simulations of the cross-correlation $F_+(t)$.
We can observe that the cross-correlation decays exponentially, 
$F_+(t)/F_+(0)=e^{-\Gamma_\mathrm{s} t}$, and its decay is faster as the coupling becomes stronger.
This numerical result completely agrees with the analytical expression~\eqref{eq:cross_analytics}, indicating that the assumption of the rare detection holds well.
As shown in the inset of Fig.~\ref{fig:cross-correlation}(a), we observe that the normalized cross-correlations $F_i(t)/F_i(0^{+})$ for different $\gamma_m^i=0.01-0.1$ collapse to the same curve.
This indicates that the strength of the quantum measurement appears only in the amplitude of the cross-correlation, being linear to $\gamma_m^i$.
The inset also confirms that $F_+(t)/\gamma_m^+=-F_-(r)/\gamma_m^->0$ holds.

Finally, let us consider the integrated cross-correlation,
\begin{align}
G_i=\int_0^\infty dt \, F_i(t).
\end{align}
The temperature dependence of the integrated cross-correlation is shown in Fig.~\ref{fig:cross-correlation}(b) for $\alpha=0.01$, 0.05, and 0.1.
The numerical simulations for different values of $\alpha$ collapse to the same curve, which indicates that the integrated cross-correlation is independent of the coupling strength $\alpha$.
These numerical simulations coincide with the analytic expression of $G_+/\Delta t$ obtained from Eq.~\eqref{eq:cross_analytics}, 
\begin{align}
\label{eq:G_analytics}
\frac{G_\pm}{\Delta t}=\pm\gamma_m^{\pm} k_\mathrm{B}\delta T~\frac{\tanh(\beta\hbar\Delta/2)}{8\mathrm{sinhc}^2(\beta\hbar\Delta)},
\end{align}
which is indicated by the solid line in Fig.~\ref{fig:cross-correlation}(b).
The reason why $G_\pm$ is independent of $\alpha$ is explained as follows.
The integrated cross-correlation can be rewritten as
\begin{align}
\frac{G_i}{\Delta t}
=E[\Delta N_i]_\mathrm{ss}Q_i^\mathrm{ex},
\label{eq:G_excess}
\end{align}
where $E[\Delta N_i]_\mathrm{ss}$ is a stationary solution of the Poisson increment and $Q_i^\mathrm{ex}$ is excess heat due to the quantum measurement defined by
\begin{align}
Q_i^\mathrm{ex}
=\int_0^\infty dt~\left[\braket{J(t)}-\braket{J}_\mathrm{ss}\right].
\end{align}
The excess heat represents a disturbance of the two-level system by the quantum measurement, and therefore it is expected to be independent of the system-bath coupling, which governs the time scale of relaxation.
In fact, we can explicitly write down excess heat as
\begin{align}
Q_\pm^\mathrm{ex}=\pm\lim_{\delta T\to0}\frac{n_L(\Delta)-n_R(\Delta)}{2}\frac{\Delta}{2}\left|\frac{\Gamma_\mathrm{a}}{\Gamma_\mathrm{s}}\right|\frac{\Gamma_\pm}{\Gamma_\mathrm{s}},
\end{align}
where $\Gamma_\mathrm{a}=\sum_r(\Gamma_{r+}-\Gamma_{r-})$ and $\Gamma_\pm=\Gamma_{L\pm}+\Gamma_{R\pm}$.
Since the rates $\Gamma_\mathrm{s}$, $\Gamma_\mathrm{a}$, and $\Gamma_\pm$ are determined by Fermi's golden rule, they are proportional to the spectral density.
Therefore excess heat is independent of the information of the system-bath coupling, i.e., the coupling strength, the cutoff function, and the type of heat bath.
Since $E[\Delta N_i]_\mathrm{ss}$ is also independent of $\alpha$, the integrated cross-correlation $G_i$ given in Eq.~(\ref{eq:G_excess}) becomes independent of the system-bath coupling.

\section{Experimental realization}

We expect that our theoretical proposal can be verified experimentally by using a platform consisting of superconducting circuits. 
The heat current through a small circuit can be measured with present technology, as demonstrated in recent experiments~\cite{Ronzani2018,Senior2020,Maillet2020,Pekola2021}.
In Ref.~\cite{Ronzani2018}, a transmon-type qubit was used as a two-level system with frequency $\Delta\approx5.30\,\mathrm{GHz}$, corresponding to $\hbar \Delta/k_\mathrm{B}\approx40.5\,\mathrm{mK}$, and the measurement was performed in the temperature range of $T\approx75-350\,\mathrm{mK}$.
Thus the condition, $\hbar\Delta\approx k_\mathrm{B}T$, required to observe the backaction discussed in this paper can be fulfilled.
To measure the integrated cross-correlation, the outcomes from continuous monitoring of the two-level system should be used as a trigger for the subsequent measurement of the integrated heat current within a finite period, which should be larger than the inverse of the decay rate $\Gamma_\mathrm{s}^{-1}$. 
This conditional heat current will deviate from the steady-state heat current without monitoring.
We emphasize that the sign of this deviation depends on which of the states $\ket{\pm}$ is used for the continuous quantum measurement on the two-level system.

\section{Summary}

We considered heat transport through a two-level system under a continuous quantum measurement onto the ground or excited state of the two-level system.
We found, in the nonselective measurement scheme, that the heat current is affected by the quantum measurement through the dephasing effect.
In addition, to observe the backaction effect directly, we calculated the cross-correlation between the measurement outcomes and the heat current.
The introduction of this quantity constitutes a first step towards evaluating the backaction of continuous quantum measurement on nonequilibrium transport phenomena in light of feasible setups.
We found that the cross-correlation depends on which of the eigenstates of the two-level system is to be measured and that the integrated cross-correlation is enhanced when the temperature becomes comparable with the energy splitting of the two-level system.
Our work, focusing on a weak system-bath coupling, represents a starting point for further research on measurement-induced transport phenomena in feasible setups.
It remains challenging to clarify new quantum many-body effects induced by quantum measurement.
We believe that our work provides a cross-cutting bridge between condensed matter physics and information theory from the viewpoint of nonequilibrium transport phenomena under continuous quantum measurement.


\section*{Acknowledgement}

Y.T. and T.K. acknowledge JSPS KAKENHI for Grants No.~JP20K03831 and No.~JP20H01827.
Y.T. and T.Y. were supported by JST Moonshot R\&D--MILLENNIA Program (Grant
No. JPMJMS2061).


\appendix

\section{Numerical calculation of $\braket{\sigma_i(t)}_\mathrm{c}$}
\label{app:sigma}

Here, we provide details on the numerical calculation of the conditional coherences and population $\braket{\sigma_i(t)}_\mathrm{c}$, which allowed us to plot Fig.~3 in the main text.

We start by expressing the reduced conditional density matrix using the Bloch vector as
\begin{align}
\tilde{\rho}_\mathrm{c}(t)
&=\frac{1}{2}\left(I+\braket{\vec{\sigma}(t)}_\mathrm{c}\cdot\vec{\sigma}\right),
\end{align}
where $\braket{\vec{\sigma}(t)}_\mathrm{c}=(\braket{\sigma_x(t)}_\mathrm{c},\braket{\sigma_y(t)}_\mathrm{c},\braket{\sigma_z(t)}_\mathrm{c})^t$ and $\vec{\sigma}=(\sigma_x,\sigma_y,\sigma_z)^t$
The stochastic master equation~(8) then reads, in the matrix representation, 
\begin{widetext}
\begin{align}
\begin{pmatrix}
1+\braket{\sigma_z'}_\mathrm{c} &
\braket{\sigma_x'}_\mathrm{c}-i\braket{\sigma_y'}_\mathrm{c} \\
\braket{\sigma_x'}_\mathrm{c}+i\braket{\sigma_y'}_\mathrm{c} & 
1-\braket{\sigma_z'}_\mathrm{c} \\
\end{pmatrix} 
&=
\begin{pmatrix}
1+\braket{\sigma_z}_\mathrm{c} & 
\braket{\sigma_x}_\mathrm{c}-i\braket{\sigma_y}_\mathrm{c} \\
\braket{\sigma_x}_\mathrm{c}+i\braket{\sigma_y}_\mathrm{c} & 
1-\braket{\sigma_z}_\mathrm{c} \\
\end{pmatrix}
-\Delta
\begin{pmatrix}
\braket{\sigma_y}_\mathrm{c} &
i\braket{\sigma_z}_\mathrm{c} \\
-i\braket{\sigma_z}_\mathrm{c} &
-\braket{\sigma_y}_\mathrm{c} \\
\end{pmatrix}\Delta t \nonumber\\
&\quad-\frac{\Gamma_\mathrm{s}}{2}
\begin{pmatrix}
\braket{\sigma_z}_\mathrm{c} &
2\braket{\sigma_x}_\mathrm{c}-i\braket{\sigma_y}_\mathrm{c} \\
2\braket{\sigma_x}_\mathrm{c}+i\braket{\sigma_y}_\mathrm{c} &
-\braket{\sigma_z}_\mathrm{c} \\
\end{pmatrix}\Delta t
-\Gamma_\mathrm{a}
\begin{pmatrix}
0 &
1 \\
1 &
0 \\
\end{pmatrix}\Delta t \nonumber\\
&\quad\mp\frac{\gamma_m^\pm}{2}
\begin{pmatrix}
\braket{\sigma_x}_\mathrm{c} &
1 \\
1 &
\braket{\sigma_x}_\mathrm{c} \\
\end{pmatrix}\Delta t
\mp\frac{\gamma_m^\pm}{2}\braket{\sigma_x}_\mathrm{c}
\begin{pmatrix}
1+\braket{\sigma_z}_\mathrm{c} &
\braket{\sigma_x}_\mathrm{c}-i\braket{\sigma_y}_\mathrm{c} \\
\braket{\sigma_x}_\mathrm{c}+i\braket{\sigma_y}_\mathrm{c} &
1-\braket{\sigma_z}_\mathrm{c} \\
\end{pmatrix}\Delta t \nonumber\\
&\quad-
\begin{pmatrix}
\braket{\sigma_z}_\mathrm{c} &
\braket{\sigma_x}_\mathrm{c}\mp1-i\braket{\sigma_y}_\mathrm{c} \\
\braket{\sigma_x}_\mathrm{c}\mp1+i\braket{\sigma_y}_\mathrm{c} &
-\braket{\sigma_z}_\mathrm{c} \\
\end{pmatrix}\Delta N_\pm,
\end{align}
\end{widetext}
where we omitted the time arguments as $\braket{\sigma_i(t)}_\mathrm{c}\to\braket{\sigma_i}_\mathrm{c}$ and $\braket{\sigma_i(t+\Delta t)}_\mathrm{c}\to\braket{\sigma_i'}_\mathrm{c}$.
Here, $\pm$ denotes the continuous quantum measurement onto an eigenstate $\ket{\pm}$ of $\sigma_x$.
Therefore the time evolution of the conditional coherences and population can be described by
\begin{widetext}
\begin{align}
\label{eq:sigmax}
\braket{\sigma_x(t+\Delta)}_\mathrm{c}
&=\braket{\sigma_x(t)}_\mathrm{c}-\left[\Gamma_\mathrm{s}\braket{\sigma_x(t)}_\mathrm{c}\pm\frac{\gamma_m^\pm}{2}\left(1-\braket{\sigma_x(t)}_\mathrm{c}^2\right)+\Gamma_\mathrm{a}\right]\Delta t-\left(\braket{\sigma_x(t)}_\mathrm{c}\mp1\right)\Delta N_\pm, \\
\label{eq:sigmay}
\braket{\sigma_y(t+\Delta t)}_\mathrm{c}
&=\braket{\sigma_y(t)}_\mathrm{c}+\left(\Delta\braket{\sigma_z(t)}_\mathrm{c}-\frac{\Gamma_\mathrm{s}}{2}\braket{\sigma_y(t)}_\mathrm{c}\pm\frac{\gamma_m^\pm}{2}\braket{\sigma_x(t)}_\mathrm{c}\braket{\sigma_y(t)}_\mathrm{c}\right)\Delta t-\braket{\sigma_y(t)}_\mathrm{c}\Delta N_\pm, \\
\label{eq:sigmaz}
\braket{\sigma_z(t+\Delta t)}_\mathrm{c}
&=\braket{\sigma_z(t)}_\mathrm{c}-\left(\Delta\braket{\sigma_y(t)}_\mathrm{c}+\frac{\Gamma_\mathrm{s}}{2}\braket{\sigma_z(t)}_\mathrm{c}\mp\frac{\gamma_m^\pm}{2}\braket{\sigma_x(t)}_\mathrm{c}\braket{\sigma_z(t)}_\mathrm{c}\right)\Delta t-\braket{\sigma_z(t)}_\mathrm{c}\Delta N_\pm.
\end{align}
\end{widetext}

Next, the Poisson variable $\Delta N_\pm$, which takes 0 or 1, is generated with the probability
\begin{align}
P[\Delta N_\pm=1]&=\frac{\gamma_m^\pm \Delta t}{2}\left(1\pm\braket{\sigma_x(t)}_\mathrm{c}\right), \\
P[\Delta N_\pm=0]&=1-P[\Delta N_\pm=1].
\end{align}
Then, we can obtain $\braket{\sigma_i(t+\Delta t)}_\mathrm{c}$ numerically by using values of $\braket{\sigma_i(t)}_\mathrm{c}$ and Eqs.~\eqref{eq:sigmax}$-$\eqref{eq:sigmaz} iteratively from the initial conditions, $\braket{\sigma_x(0)}_\mathrm{c}=\braket{\sigma_y(0)}_\mathrm{c}=0$ and $\braket{\sigma_z(0)}_\mathrm{c}=1$.

\section{Heat current under the selective measurement for $\Delta N_i=1$}
\label{app:current-DN=0}

Here, we discuss the heat current when the detected state of the two-level system is an eigen state $\ket{i=+}$ or $\ket{i=-}$ ($\Delta N_i=1$).

Using the definition of the heat current and the Heisenberg equation, the heat current is expressed as
\begin{align}
\label{eq:heat_current}
\braket{J_L(t)}_\mathrm{c}
&=-\Braket{\frac{dH_{\mathrm{B},L}}{dt}}_\mathrm{c} \nonumber\\
&=\frac{i}{2}\sum_{k}\lambda_{Lk}\hbar\omega_{Lk}\Braket{\sigma_z(t)\left[b_{Lk}(t)-b^\dagger_{Lk}(t)\right]}_\mathrm{c},
\end{align}
where $\braket{\mathcal{O}}_\mathrm{c}=\mathrm{tr}[\rho_\mathrm{c}\mathcal{O}]$.
Instantaneously after the detection at $t$, the conditional density matrix is projected to the eigenstate $\ket{i}$,
\begin{align}
\label{eq:rho_c_DeltaN=1}
\rho_\mathrm{c}(t)=\frac{P_i\rho_\mathrm{c}(t-\Delta t)P_i}{\braket{P_i(t)}_\mathrm{c}},
\end{align}
where $P_\pm=\ket{\pm}\bra{\pm}$ is the projective operator.
By plugging the conditional density matrix~\eqref{eq:rho_c_DeltaN=1} into the bracket in the heat current~\eqref{eq:heat_current}, we obtain
\begin{align}
&\Braket{\sigma_z(t)\left[b_{Lk}(t)-b^\dagger_{Lk}(t)\right]}_\mathrm{c} \nonumber\\
&=\mathrm{tr}\left[\rho_\mathrm{c}(t)\sigma_z\left(b_{Lk}-b^\dagger_{Lk}\right)\right] \nonumber\\
&=\frac{\mathrm{tr}\left[\rho_\mathrm{c}(t-\Delta t)P_i\sigma_zP_i\left(b_{Lk}-b^\dagger_{Lk}\right)\right]}{\braket{P_i(t)}_\mathrm{c}}=0,
\end{align}
where we have used the cyclic property, $[P_i,b^{(\dagger)}_{Lk}]=0$, and $P_i\sigma_zP_i=0$.
Therefore, when $\Delta N_i=1$, heat current does not flow, i.e., $\braket{J_L(t)}_\mathrm{c}=0$.

Finally, we note that the contribution of the heat current at the detection does not matter for the ensemble values in our numerical simulations at the $\Delta t\to0$ limit.

\bibliography{heat-measurement}

\begin{thebibliography}{65}%
\makeatletter
\providecommand \@ifxundefined [1]{%
 \@ifx{#1\undefined}
}%
\providecommand \@ifnum [1]{%
 \ifnum #1\expandafter \@firstoftwo
 \else \expandafter \@secondoftwo
 \fi
}%
\providecommand \@ifx [1]{%
 \ifx #1\expandafter \@firstoftwo
 \else \expandafter \@secondoftwo
 \fi
}%
\providecommand \natexlab [1]{#1}%
\providecommand \enquote  [1]{``#1''}%
\providecommand \bibnamefont  [1]{#1}%
\providecommand \bibfnamefont [1]{#1}%
\providecommand \citenamefont [1]{#1}%
\providecommand \href@noop [0]{\@secondoftwo}%
\providecommand \href [0]{\begingroup \@sanitize@url \@href}%
\providecommand \@href[1]{\@@startlink{#1}\@@href}%
\providecommand \@@href[1]{\endgroup#1\@@endlink}%
\providecommand \@sanitize@url [0]{\catcode `\\12\catcode `\$12\catcode
  `\&12\catcode `\#12\catcode `\^12\catcode `\_12\catcode `\%12\relax}%
\providecommand \@@startlink[1]{}%
\providecommand \@@endlink[0]{}%
\providecommand \url  [0]{\begingroup\@sanitize@url \@url }%
\providecommand \@url [1]{\endgroup\@href {#1}{\urlprefix }}%
\providecommand \urlprefix  [0]{URL }%
\providecommand \Eprint [0]{\href }%
\providecommand \doibase [0]{http://dx.doi.org/}%
\providecommand \selectlanguage [0]{\@gobble}%
\providecommand \bibinfo  [0]{\@secondoftwo}%
\providecommand \bibfield  [0]{\@secondoftwo}%
\providecommand \translation [1]{[#1]}%
\providecommand \BibitemOpen [0]{}%
\providecommand \bibitemStop [0]{}%
\providecommand \bibitemNoStop [0]{.\EOS\space}%
\providecommand \EOS [0]{\spacefactor3000\relax}%
\providecommand \BibitemShut  [1]{\csname bibitem#1\endcsname}%
\let\auto@bib@innerbib\@empty
\bibitem [{\citenamefont {Wiseman}\ and\ \citenamefont
  {Milburn}(2009)}]{Wiseman_text}%
  \BibitemOpen
  \bibfield  {author} {\bibinfo {author} {\bibfnamefont {H.~M.}\ \bibnamefont
  {Wiseman}}\ and\ \bibinfo {author} {\bibfnamefont {G.~J.}\ \bibnamefont
  {Milburn}},\ }\href@noop {} {\emph {\bibinfo {title} {Quantum Measurement and
  Control}}}\ (\bibinfo  {publisher} {Cambridge University Press},\ \bibinfo
  {address} {Cambridge},\ \bibinfo {year} {2009})\BibitemShut {NoStop}%
\bibitem [{\citenamefont {Li}\ \emph {et~al.}(2018)\citenamefont {Li},
  \citenamefont {Chen},\ and\ \citenamefont {Fisher}}]{Li2018}%
  \BibitemOpen
  \bibfield  {author} {\bibinfo {author} {\bibfnamefont {Y.}~\bibnamefont
  {Li}}, \bibinfo {author} {\bibfnamefont {X.}~\bibnamefont {Chen}}, \ and\
  \bibinfo {author} {\bibfnamefont {M.~P.~A.}\ \bibnamefont {Fisher}},\
  }\href@noop {} {\bibfield  {journal} {\bibinfo  {journal} {Phys. Rev. B}\
  }\textbf {\bibinfo {volume} {98}},\ \bibinfo {pages} {205136} (\bibinfo
  {year} {2018})}\BibitemShut {NoStop}%
\bibitem [{\citenamefont {Li}\ \emph {et~al.}(2019)\citenamefont {Li},
  \citenamefont {Chen},\ and\ \citenamefont {Fisher}}]{Li2019}%
  \BibitemOpen
  \bibfield  {author} {\bibinfo {author} {\bibfnamefont {Y.}~\bibnamefont
  {Li}}, \bibinfo {author} {\bibfnamefont {X.}~\bibnamefont {Chen}}, \ and\
  \bibinfo {author} {\bibfnamefont {M.~P.~A.}\ \bibnamefont {Fisher}},\
  }\href@noop {} {\bibfield  {journal} {\bibinfo  {journal} {Phys. Rev. B}\
  }\textbf {\bibinfo {volume} {100}},\ \bibinfo {pages} {134306} (\bibinfo
  {year} {2019})}\BibitemShut {NoStop}%
\bibitem [{\citenamefont {Skinner}\ \emph {et~al.}(2019)\citenamefont
  {Skinner}, \citenamefont {Ruhman},\ and\ \citenamefont
  {Nahum}}]{Skinner2019}%
  \BibitemOpen
  \bibfield  {author} {\bibinfo {author} {\bibfnamefont {B.}~\bibnamefont
  {Skinner}}, \bibinfo {author} {\bibfnamefont {J.}~\bibnamefont {Ruhman}}, \
  and\ \bibinfo {author} {\bibfnamefont {A.}~\bibnamefont {Nahum}},\
  }\href@noop {} {\bibfield  {journal} {\bibinfo  {journal} {Phys. Rev. X}\
  }\textbf {\bibinfo {volume} {9}},\ \bibinfo {pages} {031009} (\bibinfo {year}
  {2019})}\BibitemShut {NoStop}%
\bibitem [{\citenamefont {Ippoliti}\ \emph {et~al.}(2021)\citenamefont
  {Ippoliti}, \citenamefont {Gullans}, \citenamefont {Gopalakrishnan},
  \citenamefont {Huse},\ and\ \citenamefont {Khemani}}]{Ippoliti2021}%
  \BibitemOpen
  \bibfield  {author} {\bibinfo {author} {\bibfnamefont {M.}~\bibnamefont
  {Ippoliti}}, \bibinfo {author} {\bibfnamefont {M.~J.}\ \bibnamefont
  {Gullans}}, \bibinfo {author} {\bibfnamefont {S.}~\bibnamefont
  {Gopalakrishnan}}, \bibinfo {author} {\bibfnamefont {D.~A.}\ \bibnamefont
  {Huse}}, \ and\ \bibinfo {author} {\bibfnamefont {V.}~\bibnamefont
  {Khemani}},\ }\href@noop {} {\bibfield  {journal} {\bibinfo  {journal} {Phys.
  Rev. X}\ }\textbf {\bibinfo {volume} {11}},\ \bibinfo {pages} {011030}
  (\bibinfo {year} {2021})}\BibitemShut {NoStop}%
\bibitem [{\citenamefont {Minato}\ \emph {et~al.}(2022)\citenamefont {Minato},
  \citenamefont {Sugimoto}, \citenamefont {Kuwahara},\ and\ \citenamefont
  {Saito}}]{Minato2022}%
  \BibitemOpen
  \bibfield  {author} {\bibinfo {author} {\bibfnamefont {T.}~\bibnamefont
  {Minato}}, \bibinfo {author} {\bibfnamefont {K.}~\bibnamefont {Sugimoto}},
  \bibinfo {author} {\bibfnamefont {T.}~\bibnamefont {Kuwahara}}, \ and\
  \bibinfo {author} {\bibfnamefont {K.}~\bibnamefont {Saito}},\ }\href@noop {}
  {\bibfield  {journal} {\bibinfo  {journal} {Phys. Rev. Lett.}\ }\textbf
  {\bibinfo {volume} {128}},\ \bibinfo {pages} {010603} (\bibinfo {year}
  {2022})}\BibitemShut {NoStop}%
\bibitem [{\citenamefont {Lee}\ and\ \citenamefont {Chan}(2014)}]{Lee2014}%
  \BibitemOpen
  \bibfield  {author} {\bibinfo {author} {\bibfnamefont {T.~E.}\ \bibnamefont
  {Lee}}\ and\ \bibinfo {author} {\bibfnamefont {C.-K.}\ \bibnamefont {Chan}},\
  }\href@noop {} {\bibfield  {journal} {\bibinfo  {journal} {Phys. Rev. X}\
  }\textbf {\bibinfo {volume} {4}},\ \bibinfo {pages} {041001} (\bibinfo {year}
  {2014})}\BibitemShut {NoStop}%
\bibitem [{\citenamefont {Ashida}\ \emph {et~al.}(2016)\citenamefont {Ashida},
  \citenamefont {Furukawa},\ and\ \citenamefont {Ueda}}]{Ashida2016}%
  \BibitemOpen
  \bibfield  {author} {\bibinfo {author} {\bibfnamefont {Y.}~\bibnamefont
  {Ashida}}, \bibinfo {author} {\bibfnamefont {S.}~\bibnamefont {Furukawa}}, \
  and\ \bibinfo {author} {\bibfnamefont {M.}~\bibnamefont {Ueda}},\ }\href@noop
  {} {\bibfield  {journal} {\bibinfo  {journal} {Phys. Rev. A}\ }\textbf
  {\bibinfo {volume} {94}},\ \bibinfo {pages} {053615} (\bibinfo {year}
  {2016})}\BibitemShut {NoStop}%
\bibitem [{\citenamefont {Ashida}\ and\ \citenamefont
  {Ueda}(2018)}]{Ashida2018}%
  \BibitemOpen
  \bibfield  {author} {\bibinfo {author} {\bibfnamefont {Y.}~\bibnamefont
  {Ashida}}\ and\ \bibinfo {author} {\bibfnamefont {M.}~\bibnamefont {Ueda}},\
  }\href@noop {} {\bibfield  {journal} {\bibinfo  {journal} {Phys. Rev. Lett.}\
  }\textbf {\bibinfo {volume} {120}},\ \bibinfo {pages} {185301} (\bibinfo
  {year} {2018})}\BibitemShut {NoStop}%
\bibitem [{\citenamefont {Hasegawa}\ \emph {et~al.}()\citenamefont {Hasegawa},
  \citenamefont {Nakagawa},\ and\ \citenamefont {Saito}}]{Hasegawa2022}%
  \BibitemOpen
  \bibfield  {author} {\bibinfo {author} {\bibfnamefont {M.}~\bibnamefont
  {Hasegawa}}, \bibinfo {author} {\bibfnamefont {M.}~\bibnamefont {Nakagawa}},
  \ and\ \bibinfo {author} {\bibfnamefont {K.}~\bibnamefont {Saito}},\
  }\href@noop {} {\bibinfo  {journal} {arXiv:2111.07771}\ }\BibitemShut
  {NoStop}%
\bibitem [{\citenamefont {Murch}\ \emph {et~al.}(2008)\citenamefont {Murch},
  \citenamefont {Moore}, \citenamefont {Gupta},\ and\ \citenamefont
  {Stamper-Kurn}}]{Murch2008}%
  \BibitemOpen
\bibfield  {journal} {  }\bibfield  {author} {\bibinfo {author} {\bibfnamefont
  {K.~W.}\ \bibnamefont {Murch}}, \bibinfo {author} {\bibfnamefont {K.~L.}\
  \bibnamefont {Moore}}, \bibinfo {author} {\bibfnamefont {S.}~\bibnamefont
  {Gupta}}, \ and\ \bibinfo {author} {\bibfnamefont {D.~M.}\ \bibnamefont
  {Stamper-Kurn}},\ }\href@noop {} {\bibfield  {journal} {\bibinfo  {journal}
  {Nature Physics}\ }\textbf {\bibinfo {volume} {4}},\ \bibinfo {pages} {561}
  (\bibinfo {year} {2008})}\BibitemShut {NoStop}%
\bibitem [{\citenamefont {Syassen}\ \emph {et~al.}(2008)\citenamefont
  {Syassen}, \citenamefont {Bauer}, \citenamefont {Lettner}, \citenamefont
  {Volz}, \citenamefont {Dietze}, \citenamefont {García-Ripoll}, \citenamefont
  {Cirac}, \citenamefont {Rempe},\ and\ \citenamefont {Dürr}}]{Syassen2014}%
  \BibitemOpen
  \bibfield  {author} {\bibinfo {author} {\bibfnamefont {N.}~\bibnamefont
  {Syassen}}, \bibinfo {author} {\bibfnamefont {D.~M.}\ \bibnamefont {Bauer}},
  \bibinfo {author} {\bibfnamefont {M.}~\bibnamefont {Lettner}}, \bibinfo
  {author} {\bibfnamefont {T.}~\bibnamefont {Volz}}, \bibinfo {author}
  {\bibfnamefont {D.}~\bibnamefont {Dietze}}, \bibinfo {author} {\bibfnamefont
  {J.~J.}\ \bibnamefont {García-Ripoll}}, \bibinfo {author} {\bibfnamefont
  {J.~I.}\ \bibnamefont {Cirac}}, \bibinfo {author} {\bibfnamefont
  {G.}~\bibnamefont {Rempe}}, \ and\ \bibinfo {author} {\bibfnamefont
  {S.}~\bibnamefont {Dürr}},\ }\href@noop {} {\bibfield  {journal} {\bibinfo
  {journal} {Science}\ }\textbf {\bibinfo {volume} {320}},\ \bibinfo {pages}
  {1329} (\bibinfo {year} {2008})}\BibitemShut {NoStop}%
\bibitem [{\citenamefont {Zhu}\ \emph {et~al.}(2014)\citenamefont {Zhu},
  \citenamefont {Gadway}, \citenamefont {Foss-Feig}, \citenamefont
  {Schachenmayer}, \citenamefont {Wall}, \citenamefont {Hazzard}, \citenamefont
  {Yan}, \citenamefont {Moses}, \citenamefont {Covey}, \citenamefont {Jin},
  \citenamefont {Ye}, \citenamefont {Holland},\ and\ \citenamefont
  {Rey}}]{Zhu2014}%
  \BibitemOpen
  \bibfield  {author} {\bibinfo {author} {\bibfnamefont {B.}~\bibnamefont
  {Zhu}}, \bibinfo {author} {\bibfnamefont {B.}~\bibnamefont {Gadway}},
  \bibinfo {author} {\bibfnamefont {M.}~\bibnamefont {Foss-Feig}}, \bibinfo
  {author} {\bibfnamefont {J.}~\bibnamefont {Schachenmayer}}, \bibinfo {author}
  {\bibfnamefont {M.~L.}\ \bibnamefont {Wall}}, \bibinfo {author}
  {\bibfnamefont {K.~R.~A.}\ \bibnamefont {Hazzard}}, \bibinfo {author}
  {\bibfnamefont {B.}~\bibnamefont {Yan}}, \bibinfo {author} {\bibfnamefont
  {S.~A.}\ \bibnamefont {Moses}}, \bibinfo {author} {\bibfnamefont {J.~P.}\
  \bibnamefont {Covey}}, \bibinfo {author} {\bibfnamefont {D.~S.}\ \bibnamefont
  {Jin}}, \bibinfo {author} {\bibfnamefont {J.}~\bibnamefont {Ye}}, \bibinfo
  {author} {\bibfnamefont {M.}~\bibnamefont {Holland}}, \ and\ \bibinfo
  {author} {\bibfnamefont {A.~M.}\ \bibnamefont {Rey}},\ }\href@noop {}
  {\bibfield  {journal} {\bibinfo  {journal} {Phys. Rev. Lett.}\ }\textbf
  {\bibinfo {volume} {112}},\ \bibinfo {pages} {070404} (\bibinfo {year}
  {2014})}\BibitemShut {NoStop}%
\bibitem [{\citenamefont {Patil}\ \emph {et~al.}(2015)\citenamefont {Patil},
  \citenamefont {Chakram},\ and\ \citenamefont {Vengalattore}}]{Patil2015}%
  \BibitemOpen
  \bibfield  {author} {\bibinfo {author} {\bibfnamefont {Y.~S.}\ \bibnamefont
  {Patil}}, \bibinfo {author} {\bibfnamefont {S.}~\bibnamefont {Chakram}}, \
  and\ \bibinfo {author} {\bibfnamefont {M.}~\bibnamefont {Vengalattore}},\
  }\href@noop {} {\bibfield  {journal} {\bibinfo  {journal} {Phys. Rev. Lett.}\
  }\textbf {\bibinfo {volume} {115}},\ \bibinfo {pages} {140402} (\bibinfo
  {year} {2015})}\BibitemShut {NoStop}%
\bibitem [{\citenamefont {Tomita}\ \emph {et~al.}(2017)\citenamefont {Tomita},
  \citenamefont {Nakajima}, \citenamefont {Danshita}, \citenamefont {Takasu},\
  and\ \citenamefont {Takahashi}}]{Tomita2017}%
  \BibitemOpen
  \bibfield  {author} {\bibinfo {author} {\bibfnamefont {T.}~\bibnamefont
  {Tomita}}, \bibinfo {author} {\bibfnamefont {S.}~\bibnamefont {Nakajima}},
  \bibinfo {author} {\bibfnamefont {I.}~\bibnamefont {Danshita}}, \bibinfo
  {author} {\bibfnamefont {Y.}~\bibnamefont {Takasu}}, \ and\ \bibinfo {author}
  {\bibfnamefont {Y.}~\bibnamefont {Takahashi}},\ }\href@noop {} {\bibfield
  {journal} {\bibinfo  {journal} {Science Advances}\ }\textbf {\bibinfo
  {volume} {3}},\ \bibinfo {pages} {1701513} (\bibinfo {year}
  {2017})}\BibitemShut {NoStop}%
\bibitem [{\citenamefont {Bischoff}\ \emph {et~al.}(2015)\citenamefont
  {Bischoff}, \citenamefont {Eich}, \citenamefont {Zilberberg}, \citenamefont
  {R{\"o}ssler}, \citenamefont {Ihn},\ and\ \citenamefont
  {Ensslin}}]{Bischoff2015}%
  \BibitemOpen
  \bibfield  {author} {\bibinfo {author} {\bibfnamefont {D.}~\bibnamefont
  {Bischoff}}, \bibinfo {author} {\bibfnamefont {M.}~\bibnamefont {Eich}},
  \bibinfo {author} {\bibfnamefont {O.}~\bibnamefont {Zilberberg}}, \bibinfo
  {author} {\bibfnamefont {C.}~\bibnamefont {R{\"o}ssler}}, \bibinfo {author}
  {\bibfnamefont {T.}~\bibnamefont {Ihn}}, \ and\ \bibinfo {author}
  {\bibfnamefont {K.}~\bibnamefont {Ensslin}},\ }\href@noop {} {\bibfield
  {journal} {\bibinfo  {journal} {Nano Lett.}\ }\textbf {\bibinfo {volume}
  {15}},\ \bibinfo {pages} {6003} (\bibinfo {year} {2015})}\BibitemShut
  {NoStop}%
\bibitem [{\citenamefont {Hatridge}\ \emph {et~al.}(2013)\citenamefont
  {Hatridge}, \citenamefont {Shankar}, \citenamefont {Mirrahimi}, \citenamefont
  {Schackert}, \citenamefont {Geerlings}, \citenamefont {Brecht}, \citenamefont
  {Sliwa}, \citenamefont {Abdo}, \citenamefont {Frunzio}, \citenamefont
  {Girvin}, \citenamefont {Schoelkopf},\ and\ \citenamefont
  {Devoret}}]{Hatridge2013}%
  \BibitemOpen
  \bibfield  {author} {\bibinfo {author} {\bibfnamefont {M.}~\bibnamefont
  {Hatridge}}, \bibinfo {author} {\bibfnamefont {S.}~\bibnamefont {Shankar}},
  \bibinfo {author} {\bibfnamefont {M.}~\bibnamefont {Mirrahimi}}, \bibinfo
  {author} {\bibfnamefont {F.}~\bibnamefont {Schackert}}, \bibinfo {author}
  {\bibfnamefont {K.}~\bibnamefont {Geerlings}}, \bibinfo {author}
  {\bibfnamefont {T.}~\bibnamefont {Brecht}}, \bibinfo {author} {\bibfnamefont
  {K.~M.}\ \bibnamefont {Sliwa}}, \bibinfo {author} {\bibfnamefont
  {B.}~\bibnamefont {Abdo}}, \bibinfo {author} {\bibfnamefont {L.}~\bibnamefont
  {Frunzio}}, \bibinfo {author} {\bibfnamefont {S.~M.}\ \bibnamefont {Girvin}},
  \bibinfo {author} {\bibfnamefont {R.~J.}\ \bibnamefont {Schoelkopf}}, \ and\
  \bibinfo {author} {\bibfnamefont {M.~H.}\ \bibnamefont {Devoret}},\
  }\href@noop {} {\bibfield  {journal} {\bibinfo  {journal} {Science}\ }\textbf
  {\bibinfo {volume} {339}},\ \bibinfo {pages} {178} (\bibinfo {year}
  {2013})}\BibitemShut {NoStop}%
\bibitem [{\citenamefont {Groen}\ \emph {et~al.}(2013)\citenamefont {Groen},
  \citenamefont {Rist\`e}, \citenamefont {Tornberg}, \citenamefont {Cramer},
  \citenamefont {de~Groot}, \citenamefont {Picot}, \citenamefont {Johansson},\
  and\ \citenamefont {DiCarlo}}]{Groen2013}%
  \BibitemOpen
  \bibfield  {author} {\bibinfo {author} {\bibfnamefont {J.~P.}\ \bibnamefont
  {Groen}}, \bibinfo {author} {\bibfnamefont {D.}~\bibnamefont {Rist\`e}},
  \bibinfo {author} {\bibfnamefont {L.}~\bibnamefont {Tornberg}}, \bibinfo
  {author} {\bibfnamefont {J.}~\bibnamefont {Cramer}}, \bibinfo {author}
  {\bibfnamefont {P.~C.}\ \bibnamefont {de~Groot}}, \bibinfo {author}
  {\bibfnamefont {T.}~\bibnamefont {Picot}}, \bibinfo {author} {\bibfnamefont
  {G.}~\bibnamefont {Johansson}}, \ and\ \bibinfo {author} {\bibfnamefont
  {L.}~\bibnamefont {DiCarlo}},\ }\href@noop {} {\bibfield  {journal} {\bibinfo
   {journal} {Phys. Rev. Lett.}\ }\textbf {\bibinfo {volume} {111}},\ \bibinfo
  {pages} {090506} (\bibinfo {year} {2013})}\BibitemShut {NoStop}%
\bibitem [{\citenamefont {Campisi}\ \emph {et~al.}(2017)\citenamefont
  {Campisi}, \citenamefont {Pekola},\ and\ \citenamefont
  {Fazio}}]{Campisi2017}%
  \BibitemOpen
  \bibfield  {author} {\bibinfo {author} {\bibfnamefont {M.}~\bibnamefont
  {Campisi}}, \bibinfo {author} {\bibfnamefont {J.}~\bibnamefont {Pekola}}, \
  and\ \bibinfo {author} {\bibfnamefont {R.}~\bibnamefont {Fazio}},\
  }\href@noop {} {\bibfield  {journal} {\bibinfo  {journal} {New Journal of
  Physics}\ }\textbf {\bibinfo {volume} {19}},\ \bibinfo {pages} {053027}
  (\bibinfo {year} {2017})}\BibitemShut {NoStop}%
\bibitem [{\citenamefont {Bhandari}\ and\ \citenamefont
  {Jordan}(2022)}]{Bhandari2022}%
  \BibitemOpen
  \bibfield  {author} {\bibinfo {author} {\bibfnamefont {B.}~\bibnamefont
  {Bhandari}}\ and\ \bibinfo {author} {\bibfnamefont {A.~N.}\ \bibnamefont
  {Jordan}},\ }\href@noop {} {\bibfield  {journal} {\bibinfo  {journal} {Phys.
  Rev. Research}\ }\textbf {\bibinfo {volume} {4}},\ \bibinfo {pages} {033103}
  (\bibinfo {year} {2022})}\BibitemShut {NoStop}%
\bibitem [{\citenamefont {Bern\'ad}\ \emph {et~al.}(2010)\citenamefont
  {Bern\'ad}, \citenamefont {J\"a\"askel\"ainen},\ and\ \citenamefont
  {Z\"ulicke}}]{Bernad2010}%
  \BibitemOpen
  \bibfield  {author} {\bibinfo {author} {\bibfnamefont {J.~Z.}\ \bibnamefont
  {Bern\'ad}}, \bibinfo {author} {\bibfnamefont {M.}~\bibnamefont
  {J\"a\"askel\"ainen}}, \ and\ \bibinfo {author} {\bibfnamefont
  {U.}~\bibnamefont {Z\"ulicke}},\ }\href@noop {} {\bibfield  {journal}
  {\bibinfo  {journal} {Phys. Rev. B}\ }\textbf {\bibinfo {volume} {81}},\
  \bibinfo {pages} {073403} (\bibinfo {year} {2010})}\BibitemShut {NoStop}%
\bibitem [{\citenamefont {Rech}\ and\ \citenamefont
  {Kehrein}(2011)}]{Rech2011}%
  \BibitemOpen
  \bibfield  {author} {\bibinfo {author} {\bibfnamefont {J.}~\bibnamefont
  {Rech}}\ and\ \bibinfo {author} {\bibfnamefont {S.}~\bibnamefont {Kehrein}},\
  }\href@noop {} {\bibfield  {journal} {\bibinfo  {journal} {Phys. Rev. Lett.}\
  }\textbf {\bibinfo {volume} {106}},\ \bibinfo {pages} {136808} (\bibinfo
  {year} {2011})}\BibitemShut {NoStop}%
\bibitem [{\citenamefont {Meschke}\ \emph {et~al.}(2006)\citenamefont
  {Meschke}, \citenamefont {Guichard},\ and\ \citenamefont
  {Pekola}}]{Meschke2006}%
  \BibitemOpen
  \bibfield  {author} {\bibinfo {author} {\bibfnamefont {M.}~\bibnamefont
  {Meschke}}, \bibinfo {author} {\bibfnamefont {W.}~\bibnamefont {Guichard}}, \
  and\ \bibinfo {author} {\bibfnamefont {J.~P.}\ \bibnamefont {Pekola}},\
  }\href@noop {} {\bibfield  {journal} {\bibinfo  {journal} {Nature}\ }\textbf
  {\bibinfo {volume} {444}},\ \bibinfo {pages} {187} (\bibinfo {year}
  {2006})}\BibitemShut {NoStop}%
\bibitem [{\citenamefont {Timofeev}\ \emph {et~al.}(2009)\citenamefont
  {Timofeev}, \citenamefont {Helle}, \citenamefont {Meschke}, \citenamefont
  {M\"ott\"onen},\ and\ \citenamefont {Pekola}}]{Timofeev2009}%
  \BibitemOpen
  \bibfield  {author} {\bibinfo {author} {\bibfnamefont {A.~V.}\ \bibnamefont
  {Timofeev}}, \bibinfo {author} {\bibfnamefont {M.}~\bibnamefont {Helle}},
  \bibinfo {author} {\bibfnamefont {M.}~\bibnamefont {Meschke}}, \bibinfo
  {author} {\bibfnamefont {M.}~\bibnamefont {M\"ott\"onen}}, \ and\ \bibinfo
  {author} {\bibfnamefont {J.~P.}\ \bibnamefont {Pekola}},\ }\href@noop {}
  {\bibfield  {journal} {\bibinfo  {journal} {Phys. Rev. Lett.}\ }\textbf
  {\bibinfo {volume} {102}},\ \bibinfo {pages} {200801} (\bibinfo {year}
  {2009})}\BibitemShut {NoStop}%
\bibitem [{\citenamefont {Partanen}\ \emph {et~al.}(2016)\citenamefont
  {Partanen}, \citenamefont {Tan}, \citenamefont {Govenius}, \citenamefont
  {Lake}, \citenamefont {M{\"a}kel{\"a}}, \citenamefont {Tanttu},\ and\
  \citenamefont {M{\"o}tt{\"o}nen}}]{Partanen2016}%
  \BibitemOpen
  \bibfield  {author} {\bibinfo {author} {\bibfnamefont {M.}~\bibnamefont
  {Partanen}}, \bibinfo {author} {\bibfnamefont {K.~Y.}\ \bibnamefont {Tan}},
  \bibinfo {author} {\bibfnamefont {J.}~\bibnamefont {Govenius}}, \bibinfo
  {author} {\bibfnamefont {R.~E.}\ \bibnamefont {Lake}}, \bibinfo {author}
  {\bibfnamefont {M.~K.}\ \bibnamefont {M{\"a}kel{\"a}}}, \bibinfo {author}
  {\bibfnamefont {T.}~\bibnamefont {Tanttu}}, \ and\ \bibinfo {author}
  {\bibfnamefont {M.}~\bibnamefont {M{\"o}tt{\"o}nen}},\ }\href@noop {}
  {\bibfield  {journal} {\bibinfo  {journal} {Nature Physics}\ }\textbf
  {\bibinfo {volume} {12}},\ \bibinfo {pages} {460} (\bibinfo {year}
  {2016})}\BibitemShut {NoStop}%
\bibitem [{\citenamefont {Ronzani}\ \emph {et~al.}(2018)\citenamefont
  {Ronzani}, \citenamefont {Karimi}, \citenamefont {Senior}, \citenamefont
  {Chang}, \citenamefont {Peltonen}, \citenamefont {Chen},\ and\ \citenamefont
  {Pekola}}]{Ronzani2018}%
  \BibitemOpen
  \bibfield  {author} {\bibinfo {author} {\bibfnamefont {A.}~\bibnamefont
  {Ronzani}}, \bibinfo {author} {\bibfnamefont {B.}~\bibnamefont {Karimi}},
  \bibinfo {author} {\bibfnamefont {J.}~\bibnamefont {Senior}}, \bibinfo
  {author} {\bibfnamefont {Y.-C.}\ \bibnamefont {Chang}}, \bibinfo {author}
  {\bibfnamefont {J.~T.}\ \bibnamefont {Peltonen}}, \bibinfo {author}
  {\bibfnamefont {C.}~\bibnamefont {Chen}}, \ and\ \bibinfo {author}
  {\bibfnamefont {J.~P.}\ \bibnamefont {Pekola}},\ }\href@noop {} {\bibfield
  {journal} {\bibinfo  {journal} {Nat. Phys.}\ }\textbf {\bibinfo {volume}
  {14}},\ \bibinfo {pages} {991} (\bibinfo {year} {2018})}\BibitemShut
  {NoStop}%
\bibitem [{\citenamefont {Senior}\ \emph {et~al.}(2020)\citenamefont {Senior},
  \citenamefont {Gubaydullin}, \citenamefont {Karimi}, \citenamefont
  {Peltonen}, \citenamefont {Ankerhold},\ and\ \citenamefont
  {Pekola}}]{Senior2020}%
  \BibitemOpen
  \bibfield  {author} {\bibinfo {author} {\bibfnamefont {J.}~\bibnamefont
  {Senior}}, \bibinfo {author} {\bibfnamefont {A.}~\bibnamefont {Gubaydullin}},
  \bibinfo {author} {\bibfnamefont {B.}~\bibnamefont {Karimi}}, \bibinfo
  {author} {\bibfnamefont {J.~T.}\ \bibnamefont {Peltonen}}, \bibinfo {author}
  {\bibfnamefont {J.}~\bibnamefont {Ankerhold}}, \ and\ \bibinfo {author}
  {\bibfnamefont {J.~P.}\ \bibnamefont {Pekola}},\ }\href@noop {} {\bibfield
  {journal} {\bibinfo  {journal} {Communications Physics}\ }\textbf {\bibinfo
  {volume} {3}},\ \bibinfo {pages} {40} (\bibinfo {year} {2020})}\BibitemShut
  {NoStop}%
\bibitem [{\citenamefont {Maillet}\ \emph {et~al.}(2020)\citenamefont
  {Maillet}, \citenamefont {Subero}, \citenamefont {Peltonen}, \citenamefont
  {Golubev},\ and\ \citenamefont {Pekola}}]{Maillet2020}%
  \BibitemOpen
  \bibfield  {author} {\bibinfo {author} {\bibfnamefont {O.}~\bibnamefont
  {Maillet}}, \bibinfo {author} {\bibfnamefont {D.}~\bibnamefont {Subero}},
  \bibinfo {author} {\bibfnamefont {J.~T.}\ \bibnamefont {Peltonen}}, \bibinfo
  {author} {\bibfnamefont {D.~S.}\ \bibnamefont {Golubev}}, \ and\ \bibinfo
  {author} {\bibfnamefont {J.~P.}\ \bibnamefont {Pekola}},\ }\href@noop {}
  {\bibfield  {journal} {\bibinfo  {journal} {Nature Communications}\ }\textbf
  {\bibinfo {volume} {11}},\ \bibinfo {pages} {4326} (\bibinfo {year}
  {2020})}\BibitemShut {NoStop}%
\bibitem [{\citenamefont {Pekola}\ and\ \citenamefont
  {Karimi}(2021)}]{Pekola2021}%
  \BibitemOpen
  \bibfield  {author} {\bibinfo {author} {\bibfnamefont {J.~P.}\ \bibnamefont
  {Pekola}}\ and\ \bibinfo {author} {\bibfnamefont {B.}~\bibnamefont
  {Karimi}},\ }\href@noop {} {\bibfield  {journal} {\bibinfo  {journal} {Rev.
  Mod. Phys.}\ }\textbf {\bibinfo {volume} {93}},\ \bibinfo {pages} {041001}
  (\bibinfo {year} {2021})}\BibitemShut {NoStop}%
\bibitem [{\citenamefont {Leggett}\ \emph {et~al.}(1987)\citenamefont
  {Leggett}, \citenamefont {Chakravarty}, \citenamefont {Dorsey}, \citenamefont
  {Fisher}, \citenamefont {Garg},\ and\ \citenamefont {Zwerger}}]{Leggett1987}%
  \BibitemOpen
  \bibfield  {author} {\bibinfo {author} {\bibfnamefont {A.~J.}\ \bibnamefont
  {Leggett}}, \bibinfo {author} {\bibfnamefont {S.}~\bibnamefont
  {Chakravarty}}, \bibinfo {author} {\bibfnamefont {A.~T.}\ \bibnamefont
  {Dorsey}}, \bibinfo {author} {\bibfnamefont {M.~P.~A.}\ \bibnamefont
  {Fisher}}, \bibinfo {author} {\bibfnamefont {A.}~\bibnamefont {Garg}}, \ and\
  \bibinfo {author} {\bibfnamefont {W.}~\bibnamefont {Zwerger}},\ }\href
  {\doibase 10.1103/RevModPhys.59.1} {\bibfield  {journal} {\bibinfo  {journal}
  {Rev. Mod. Phys.}\ }\textbf {\bibinfo {volume} {59}},\ \bibinfo {pages} {1}
  (\bibinfo {year} {1987})}\BibitemShut {NoStop}%
\bibitem [{\citenamefont {Weiss}(2012)}]{Weiss_text}%
  \BibitemOpen
  \bibfield  {author} {\bibinfo {author} {\bibfnamefont {U.}~\bibnamefont
  {Weiss}},\ }\href@noop {} {\emph {\bibinfo {title} {{Quantum Dissipative
  Systems.}}}}\ (\bibinfo  {publisher} {World Scientific},\ \bibinfo {address}
  {Singapore},\ \bibinfo {year} {2012})\BibitemShut {NoStop}%
\bibitem [{\citenamefont {Hewson}(1993)}]{Hewson_text}%
  \BibitemOpen
  \bibfield  {author} {\bibinfo {author} {\bibfnamefont {A.}~\bibnamefont
  {Hewson}},\ }\href@noop {} {\emph {\bibinfo {title} {The Kondo Problem to
  Heavy Fermions}}}\ (\bibinfo  {publisher} {Cambridge University Press},\
  \bibinfo {address} {New York},\ \bibinfo {year} {1993})\BibitemShut {NoStop}%
\bibitem [{\citenamefont {Guinea}\ \emph {et~al.}(1985)\citenamefont {Guinea},
  \citenamefont {Hakim},\ and\ \citenamefont {Muramatsu}}]{Guinea1985}%
  \BibitemOpen
  \bibfield  {author} {\bibinfo {author} {\bibfnamefont {F.}~\bibnamefont
  {Guinea}}, \bibinfo {author} {\bibfnamefont {V.}~\bibnamefont {Hakim}}, \
  and\ \bibinfo {author} {\bibfnamefont {A.}~\bibnamefont {Muramatsu}},\
  }\href@noop {} {\bibfield  {journal} {\bibinfo  {journal} {Phys. Rev. B}\
  }\textbf {\bibinfo {volume} {32}},\ \bibinfo {pages} {4410} (\bibinfo {year}
  {1985})}\BibitemShut {NoStop}%
\bibitem [{\citenamefont {Le~Hur}(2012)}]{LeHur2012}%
  \BibitemOpen
  \bibfield  {author} {\bibinfo {author} {\bibfnamefont {K.}~\bibnamefont
  {Le~Hur}},\ }\href@noop {} {\bibfield  {journal} {\bibinfo  {journal} {Phys.
  Rev. B}\ }\textbf {\bibinfo {volume} {85}},\ \bibinfo {pages} {140506(R)}
  (\bibinfo {year} {2012})}\BibitemShut {NoStop}%
\bibitem [{\citenamefont {Saito}\ and\ \citenamefont {Kato}(2013)}]{Saito2013}%
  \BibitemOpen
  \bibfield  {author} {\bibinfo {author} {\bibfnamefont {K.}~\bibnamefont
  {Saito}}\ and\ \bibinfo {author} {\bibfnamefont {T.}~\bibnamefont {Kato}},\
  }\href@noop {} {\bibfield  {journal} {\bibinfo  {journal} {Phys. Rev. Lett.}\
  }\textbf {\bibinfo {volume} {111}},\ \bibinfo {pages} {214301} (\bibinfo
  {year} {2013})}\BibitemShut {NoStop}%
\bibitem [{\citenamefont {Yamamoto}\ \emph {et~al.}(2018)\citenamefont
  {Yamamoto}, \citenamefont {Kato}, \citenamefont {Kato},\ and\ \citenamefont
  {Saito}}]{Yamamoto2018NJP}%
  \BibitemOpen
  \bibfield  {author} {\bibinfo {author} {\bibfnamefont {T.}~\bibnamefont
  {Yamamoto}}, \bibinfo {author} {\bibfnamefont {M.}~\bibnamefont {Kato}},
  \bibinfo {author} {\bibfnamefont {T.}~\bibnamefont {Kato}}, \ and\ \bibinfo
  {author} {\bibfnamefont {K.}~\bibnamefont {Saito}},\ }\href@noop {}
  {\bibfield  {journal} {\bibinfo  {journal} {New J. Phys.}\ }\textbf {\bibinfo
  {volume} {20}},\ \bibinfo {pages} {093014} (\bibinfo {year}
  {2018})}\BibitemShut {NoStop}%
\bibitem [{\citenamefont {Anderson}\ and\ \citenamefont
  {Yuval}(1971)}]{Anderson1971}%
  \BibitemOpen
  \bibfield  {author} {\bibinfo {author} {\bibfnamefont {P.~W.}\ \bibnamefont
  {Anderson}}\ and\ \bibinfo {author} {\bibfnamefont {G.}~\bibnamefont
  {Yuval}},\ }\href@noop {} {\bibfield  {journal} {\bibinfo  {journal} {J.
  Phys. C: Solid State Phys.}\ }\textbf {\bibinfo {volume} {4}},\ \bibinfo
  {pages} {607} (\bibinfo {year} {1971})}\BibitemShut {NoStop}%
\bibitem [{\citenamefont {Kosterlitz}(1976)}]{Kosterlitz1976}%
  \BibitemOpen
  \bibfield  {author} {\bibinfo {author} {\bibfnamefont {J.~M.}\ \bibnamefont
  {Kosterlitz}},\ }\href@noop {} {\bibfield  {journal} {\bibinfo  {journal}
  {Phys. Rev. Lett.}\ }\textbf {\bibinfo {volume} {37}},\ \bibinfo {pages}
  {1577} (\bibinfo {year} {1976})}\BibitemShut {NoStop}%
\bibitem [{\citenamefont {De~Filippis}\ \emph {et~al.}(2022)\citenamefont
  {De~Filippis}, \citenamefont {de~Candia}, \citenamefont {Di~Bello},
  \citenamefont {Perroni}, \citenamefont {Cangemi}, \citenamefont {Nocera},
  \citenamefont {Sassetti}, \citenamefont {Fazio},\ and\ \citenamefont
  {Cataudella}}]{Filippis2022}%
  \BibitemOpen
  \bibfield  {author} {\bibinfo {author} {\bibfnamefont {G.}~\bibnamefont
  {De~Filippis}}, \bibinfo {author} {\bibfnamefont {A.}~\bibnamefont
  {de~Candia}}, \bibinfo {author} {\bibfnamefont {G.}~\bibnamefont {Di~Bello}},
  \bibinfo {author} {\bibfnamefont {C.~A.}\ \bibnamefont {Perroni}}, \bibinfo
  {author} {\bibfnamefont {L.~M.}\ \bibnamefont {Cangemi}}, \bibinfo {author}
  {\bibfnamefont {A.}~\bibnamefont {Nocera}}, \bibinfo {author} {\bibfnamefont
  {M.}~\bibnamefont {Sassetti}}, \bibinfo {author} {\bibfnamefont
  {R.}~\bibnamefont {Fazio}}, \ and\ \bibinfo {author} {\bibfnamefont
  {V.}~\bibnamefont {Cataudella}},\ }\href@noop {} {\bibfield  {journal}
  {\bibinfo  {journal} {arXiv:2205.11555}\ } (\bibinfo {year}
  {2022})}\BibitemShut {NoStop}%
\bibitem [{\citenamefont {Kehrein}\ \emph {et~al.}(1995)\citenamefont
  {Kehrein}, \citenamefont {Mielke},\ and\ \citenamefont {Neu}}]{Kehrein1995}%
  \BibitemOpen
  \bibfield  {author} {\bibinfo {author} {\bibfnamefont {S.~K.}\ \bibnamefont
  {Kehrein}}, \bibinfo {author} {\bibfnamefont {A.}~\bibnamefont {Mielke}}, \
  and\ \bibinfo {author} {\bibfnamefont {P.}~\bibnamefont {Neu}},\ }\href@noop
  {} {\bibfield  {journal} {\bibinfo  {journal} {Z. Phys. B Condensed Matter}\
  }\textbf {\bibinfo {volume} {99}},\ \bibinfo {pages} {269} (\bibinfo {year}
  {1995})}\BibitemShut {NoStop}%
\bibitem [{\citenamefont {Kehrein}\ and\ \citenamefont
  {Mielke}(1996)}]{Kehrein1996}%
  \BibitemOpen
  \bibfield  {author} {\bibinfo {author} {\bibfnamefont {S.~K.}\ \bibnamefont
  {Kehrein}}\ and\ \bibinfo {author} {\bibfnamefont {A.}~\bibnamefont
  {Mielke}},\ }\href@noop {} {\bibfield  {journal} {\bibinfo  {journal} {Phys.
  Lett. A}\ }\textbf {\bibinfo {volume} {219}},\ \bibinfo {pages} {313}
  (\bibinfo {year} {1996})}\BibitemShut {NoStop}%
\bibitem [{\citenamefont {Bulla}\ \emph {et~al.}(2003)\citenamefont {Bulla},
  \citenamefont {Tong},\ and\ \citenamefont {Vojta}}]{Bulla2003}%
  \BibitemOpen
  \bibfield  {author} {\bibinfo {author} {\bibfnamefont {R.}~\bibnamefont
  {Bulla}}, \bibinfo {author} {\bibfnamefont {N.-H.}\ \bibnamefont {Tong}}, \
  and\ \bibinfo {author} {\bibfnamefont {M.}~\bibnamefont {Vojta}},\
  }\href@noop {} {\bibfield  {journal} {\bibinfo  {journal} {Phys. Rev. Lett.}\
  }\textbf {\bibinfo {volume} {91}},\ \bibinfo {pages} {170601} (\bibinfo
  {year} {2003})}\BibitemShut {NoStop}%
\bibitem [{\citenamefont {Winter}\ \emph {et~al.}(2009)\citenamefont {Winter},
  \citenamefont {Rieger}, \citenamefont {Vojta},\ and\ \citenamefont
  {Bulla}}]{Winter2009}%
  \BibitemOpen
  \bibfield  {author} {\bibinfo {author} {\bibfnamefont {A.}~\bibnamefont
  {Winter}}, \bibinfo {author} {\bibfnamefont {H.}~\bibnamefont {Rieger}},
  \bibinfo {author} {\bibfnamefont {M.}~\bibnamefont {Vojta}}, \ and\ \bibinfo
  {author} {\bibfnamefont {R.}~\bibnamefont {Bulla}},\ }\href@noop {}
  {\bibfield  {journal} {\bibinfo  {journal} {Phys. Rev. Lett.}\ }\textbf
  {\bibinfo {volume} {102}},\ \bibinfo {pages} {030601} (\bibinfo {year}
  {2009})}\BibitemShut {NoStop}%
\bibitem [{\citenamefont {Yamamoto}\ and\ \citenamefont
  {Kato}(2018)}]{Yamamoto2018PRB}%
  \BibitemOpen
  \bibfield  {author} {\bibinfo {author} {\bibfnamefont {T.}~\bibnamefont
  {Yamamoto}}\ and\ \bibinfo {author} {\bibfnamefont {T.}~\bibnamefont
  {Kato}},\ }\href@noop {} {\bibfield  {journal} {\bibinfo  {journal} {Phys.
  Rev. B}\ }\textbf {\bibinfo {volume} {98}},\ \bibinfo {pages} {245412}
  (\bibinfo {year} {2018})}\BibitemShut {NoStop}%
\bibitem [{\citenamefont {Pekola}(2015)}]{Pekola2015}%
  \BibitemOpen
  \bibfield  {author} {\bibinfo {author} {\bibfnamefont {J.~P.}\ \bibnamefont
  {Pekola}},\ }\href@noop {} {\bibfield  {journal} {\bibinfo  {journal} {Nature
  Physics}\ }\textbf {\bibinfo {volume} {11}},\ \bibinfo {pages} {118}
  (\bibinfo {year} {2015})}\BibitemShut {NoStop}%
\bibitem [{\citenamefont {Szilard}(1929)}]{Szilard1929}%
  \BibitemOpen
  \bibfield  {author} {\bibinfo {author} {\bibfnamefont {L.}~\bibnamefont
  {Szilard}},\ }\href@noop {} {\bibfield  {journal} {\bibinfo  {journal} {Z.
  Phys.}\ }\textbf {\bibinfo {volume} {53}},\ \bibinfo {pages} {840} (\bibinfo
  {year} {1929})}\BibitemShut {NoStop}%
\bibitem [{\citenamefont {Leff}\ and\ \citenamefont {Rex}(2003)}]{Leff_text}%
  \BibitemOpen
  \bibinfo {editor} {\bibfnamefont {H.~S.}\ \bibnamefont {Leff}}\ and\ \bibinfo
  {editor} {\bibfnamefont {A.~F.}\ \bibnamefont {Rex}},\ eds.,\ \href@noop {}
  {\emph {\bibinfo {title} {Maxwell’s Demon 2: Entropy, Classical and Quantum
  Information, Computing}}}\ (\bibinfo  {publisher} {IOP Publishing},\ \bibinfo
  {address} {Bristol},\ \bibinfo {year} {2003})\BibitemShut {NoStop}%
\bibitem [{\citenamefont {Sagawa}\ and\ \citenamefont
  {Ueda}(2008)}]{Sagawa2008}%
  \BibitemOpen
  \bibfield  {author} {\bibinfo {author} {\bibfnamefont {T.}~\bibnamefont
  {Sagawa}}\ and\ \bibinfo {author} {\bibfnamefont {M.}~\bibnamefont {Ueda}},\
  }\href@noop {} {\bibfield  {journal} {\bibinfo  {journal} {Phys. Rev. Lett.}\
  }\textbf {\bibinfo {volume} {100}},\ \bibinfo {pages} {080403} (\bibinfo
  {year} {2008})}\BibitemShut {NoStop}%
\bibitem [{\citenamefont {Sagawa}\ and\ \citenamefont
  {Ueda}(2010)}]{Sagawa2010}%
  \BibitemOpen
  \bibfield  {author} {\bibinfo {author} {\bibfnamefont {T.}~\bibnamefont
  {Sagawa}}\ and\ \bibinfo {author} {\bibfnamefont {M.}~\bibnamefont {Ueda}},\
  }\href@noop {} {\bibfield  {journal} {\bibinfo  {journal} {Phys. Rev. Lett.}\
  }\textbf {\bibinfo {volume} {104}},\ \bibinfo {pages} {090602} (\bibinfo
  {year} {2010})}\BibitemShut {NoStop}%
\bibitem [{\citenamefont {Koski}\ \emph
  {et~al.}(2014{\natexlab{a}})\citenamefont {Koski}, \citenamefont {Maisi},
  \citenamefont {Pekola},\ and\ \citenamefont {Averin}}]{Koski2014}%
  \BibitemOpen
  \bibfield  {author} {\bibinfo {author} {\bibfnamefont {J.~V.}\ \bibnamefont
  {Koski}}, \bibinfo {author} {\bibfnamefont {V.~F.}\ \bibnamefont {Maisi}},
  \bibinfo {author} {\bibfnamefont {J.~P.}\ \bibnamefont {Pekola}}, \ and\
  \bibinfo {author} {\bibfnamefont {D.~V.}\ \bibnamefont {Averin}},\
  }\href@noop {} {\bibfield  {journal} {\bibinfo  {journal} {PNAS}\ }\textbf
  {\bibinfo {volume} {111}},\ \bibinfo {pages} {13786} (\bibinfo {year}
  {2014}{\natexlab{a}})}\BibitemShut {NoStop}%
\bibitem [{\citenamefont {Koski}\ \emph
  {et~al.}(2014{\natexlab{b}})\citenamefont {Koski}, \citenamefont {Maisi},
  \citenamefont {Sagawa},\ and\ \citenamefont {Pekola}}]{Koski2014PRL}%
  \BibitemOpen
  \bibfield  {author} {\bibinfo {author} {\bibfnamefont {J.~V.}\ \bibnamefont
  {Koski}}, \bibinfo {author} {\bibfnamefont {V.~F.}\ \bibnamefont {Maisi}},
  \bibinfo {author} {\bibfnamefont {T.}~\bibnamefont {Sagawa}}, \ and\ \bibinfo
  {author} {\bibfnamefont {J.~P.}\ \bibnamefont {Pekola}},\ }\href@noop {}
  {\bibfield  {journal} {\bibinfo  {journal} {Phys. Rev. Lett.}\ }\textbf
  {\bibinfo {volume} {113}},\ \bibinfo {pages} {030601} (\bibinfo {year}
  {2014}{\natexlab{b}})}\BibitemShut {NoStop}%
\bibitem [{\citenamefont {Carmichael}(1993)}]{Carmichael_text}%
  \BibitemOpen
  \bibfield  {author} {\bibinfo {author} {\bibfnamefont {H.}~\bibnamefont
  {Carmichael}},\ }\href@noop {} {\emph {\bibinfo {title} {An Open System
  Approach to Quantum Optics}}}\ (\bibinfo  {publisher} {Springer},\ \bibinfo
  {address} {Berlin},\ \bibinfo {year} {1993})\BibitemShut {NoStop}%
\bibitem [{\citenamefont {Breuer}\ and\ \citenamefont
  {Petruccione}(2007)}]{Breuer_text}%
  \BibitemOpen
  \bibfield  {author} {\bibinfo {author} {\bibfnamefont {H.}~\bibnamefont
  {Breuer}}\ and\ \bibinfo {author} {\bibfnamefont {F.}~\bibnamefont
  {Petruccione}},\ }\href@noop {} {\emph {\bibinfo {title} {The Theory of Open
  Quantum Systems}}}\ (\bibinfo  {publisher} {Oxford University Press},\
  \bibinfo {address} {New York},\ \bibinfo {year} {2007})\BibitemShut {NoStop}%
\bibitem [{\citenamefont {Gambetta}\ \emph {et~al.}(2006)\citenamefont
  {Gambetta}, \citenamefont {Blais}, \citenamefont {Schuster}, \citenamefont
  {Wallraff}, \citenamefont {Frunzio}, \citenamefont {Majer}, \citenamefont
  {Devoret}, \citenamefont {Girvin},\ and\ \citenamefont
  {Schoelkopf}}]{Gambetta2006}%
  \BibitemOpen
  \bibfield  {author} {\bibinfo {author} {\bibfnamefont {J.}~\bibnamefont
  {Gambetta}}, \bibinfo {author} {\bibfnamefont {A.}~\bibnamefont {Blais}},
  \bibinfo {author} {\bibfnamefont {D.~I.}\ \bibnamefont {Schuster}}, \bibinfo
  {author} {\bibfnamefont {A.}~\bibnamefont {Wallraff}}, \bibinfo {author}
  {\bibfnamefont {L.}~\bibnamefont {Frunzio}}, \bibinfo {author} {\bibfnamefont
  {J.}~\bibnamefont {Majer}}, \bibinfo {author} {\bibfnamefont {M.~H.}\
  \bibnamefont {Devoret}}, \bibinfo {author} {\bibfnamefont {S.~M.}\
  \bibnamefont {Girvin}}, \ and\ \bibinfo {author} {\bibfnamefont {R.~J.}\
  \bibnamefont {Schoelkopf}},\ }\href@noop {} {\bibfield  {journal} {\bibinfo
  {journal} {Phys. Rev. A}\ }\textbf {\bibinfo {volume} {74}},\ \bibinfo
  {pages} {042318} (\bibinfo {year} {2006})}\BibitemShut {NoStop}%
\bibitem [{\citenamefont {Iyoda}\ \emph {et~al.}(2013)\citenamefont {Iyoda},
  \citenamefont {Kato}, \citenamefont {Aoki}, \citenamefont {Edamatsu},\ and\
  \citenamefont {Koshino}}]{Iyoda2013}%
  \BibitemOpen
  \bibfield  {author} {\bibinfo {author} {\bibfnamefont {E.}~\bibnamefont
  {Iyoda}}, \bibinfo {author} {\bibfnamefont {T.}~\bibnamefont {Kato}},
  \bibinfo {author} {\bibfnamefont {T.}~\bibnamefont {Aoki}}, \bibinfo {author}
  {\bibfnamefont {K.}~\bibnamefont {Edamatsu}}, \ and\ \bibinfo {author}
  {\bibfnamefont {K.}~\bibnamefont {Koshino}},\ }\href@noop {} {\bibfield
  {journal} {\bibinfo  {journal} {J. Phys. Soc. Jpn.}\ }\textbf {\bibinfo
  {volume} {82}},\ \bibinfo {pages} {014301} (\bibinfo {year}
  {2013})}\BibitemShut {NoStop}%
\bibitem [{\citenamefont {Ojanen}\ and\ \citenamefont
  {Jauho}(2008)}]{Ojanen2008}%
  \BibitemOpen
  \bibfield  {author} {\bibinfo {author} {\bibfnamefont {T.}~\bibnamefont
  {Ojanen}}\ and\ \bibinfo {author} {\bibfnamefont {A.-P.}\ \bibnamefont
  {Jauho}},\ }\href@noop {} {\bibfield  {journal} {\bibinfo  {journal} {Phys.
  Rev. Lett.}\ }\textbf {\bibinfo {volume} {100}},\ \bibinfo {pages} {155902}
  (\bibinfo {year} {2008})}\BibitemShut {NoStop}%
\bibitem [{\citenamefont {Saito}(2008)}]{Saito2008}%
  \BibitemOpen
  \bibfield  {author} {\bibinfo {author} {\bibfnamefont {K.}~\bibnamefont
  {Saito}},\ }\href@noop {} {\bibfield  {journal} {\bibinfo  {journal}
  {Europhys. Lett.}\ }\textbf {\bibinfo {volume} {83}},\ \bibinfo {pages}
  {50006} (\bibinfo {year} {2008})}\BibitemShut {NoStop}%
\bibitem [{Note1()}]{Note1}%
  \BibitemOpen
  \bibinfo {note} {Using the quantum regression theorem~\cite {Breuer_text},
  the dynamics of the correlation function $C_{zz}(t)=\mathinner {\langle
  {\sigma _z(t)\sigma _z(0)}\rangle }$ are determined by the differential
  equation, $\protect \ddot {C}_{zz}+2\protect \tilde {\Gamma }_\protect
  \mathrm {s}\protect \dot {C}_{zz}+(\Delta ^2+\protect \tilde {\Gamma
  }_\protect \mathrm {s}^2)C_{zz}=0$, and its solution is given by
  $C_{zz}(t)=e^{-\protect \tilde {\Gamma }_\protect \mathrm {s}t}[\cos (\Delta
  t)+(\protect \tilde {\Gamma }_\protect \mathrm {s}/\Delta )\sin (\Delta t)]$
  under the initial conditions $C_{zz}(0)=1$ and $\protect \dot
  {C}_{zz}(0)=0$}\BibitemShut {NoStop}%
\bibitem [{Note2()}]{Note2}%
  \BibitemOpen
  \bibinfo {note} {The heat transport process discussed in this paper is called
  a sequential tunneling process, which can describe heat transport for
  $k_\protect \mathrm {B}T\sim \hbar \Delta $. However, at lower temperatures,
  $k_\protect \mathrm {B}T\lesssim 0.1\hbar \Delta $, the sequential tunneling
  process is suppressed, and thus the co-tunneling process becomes dominant, in
  which the higher-order interaction between the two-level system and the heat
  baths plays the important role~\cite
  {Ruokola2011,Yamamoto2018NJP}.}\BibitemShut {Stop}%
\bibitem [{\citenamefont {Segal}\ and\ \citenamefont
  {Nitzan}(2005)}]{Segal2005}%
  \BibitemOpen
  \bibfield  {author} {\bibinfo {author} {\bibfnamefont {D.}~\bibnamefont
  {Segal}}\ and\ \bibinfo {author} {\bibfnamefont {A.}~\bibnamefont {Nitzan}},\
  }\href@noop {} {\bibfield  {journal} {\bibinfo  {journal} {J. Chem. Phys.}\
  }\textbf {\bibinfo {volume} {122}},\ \bibinfo {pages} {194704} (\bibinfo
  {year} {2005})}\BibitemShut {NoStop}%
\bibitem [{\citenamefont {Wiseman}\ and\ \citenamefont
  {Milburn}(1993)}]{Wiseman1993}%
  \BibitemOpen
  \bibfield  {author} {\bibinfo {author} {\bibfnamefont {H.~M.}\ \bibnamefont
  {Wiseman}}\ and\ \bibinfo {author} {\bibfnamefont {G.~J.}\ \bibnamefont
  {Milburn}},\ }\href@noop {} {\bibfield  {journal} {\bibinfo  {journal} {Phys.
  Rev. A}\ }\textbf {\bibinfo {volume} {47}},\ \bibinfo {pages} {1652}
  (\bibinfo {year} {1993})}\BibitemShut {NoStop}%
\bibitem [{Note3()}]{Note3}%
  \BibitemOpen
  \bibinfo {note} {In our simulation, the heat current under the condition,
  $\Delta N_i=1$, can be defined arbitrarily, because the number of time steps
  in this condition is much less than the number in $\Delta N_i=0$ in the limit
  $\Delta t\to 0$. See Appendix~\ref {app:current-DN=0} for
  details.}\BibitemShut {Stop}%
\bibitem [{Note4()}]{Note4}%
  \BibitemOpen
  \bibinfo {note} {We can reproduce this expression for the heat current at
  $t\to \infty $ by plugging $S(\omega )\approx \pi \delta (\omega -\Delta )$
  into Eq.~\protect \eqref {eq:current_non-selective}}\BibitemShut {NoStop}%
\bibitem [{Note5()}]{Note5}%
  \BibitemOpen
  \bibinfo {note} {The stationary solution of $\mathinner {\langle {\sigma
  _x}\rangle }$ is calculated by solving the differential equation $\mathinner
  {\langle {\protect \dot {\sigma }_x}\rangle }=-\Gamma _\protect \mathrm
  {s}\mathinner {\langle {\sigma _x}\rangle }-\Gamma _\protect \mathrm {a}$
  under the steady-state condition $\mathinner {\langle {\protect \dot {\sigma
  }_x}\rangle }=0$.}\BibitemShut {Stop}%
\bibitem [{\citenamefont {Ruokola}\ and\ \citenamefont
  {Ojanen}(2011)}]{Ruokola2011}%
  \BibitemOpen
  \bibfield  {author} {\bibinfo {author} {\bibfnamefont {T.}~\bibnamefont
  {Ruokola}}\ and\ \bibinfo {author} {\bibfnamefont {T.}~\bibnamefont
  {Ojanen}},\ }\href@noop {} {\bibfield  {journal} {\bibinfo  {journal} {Phys.
  Rev. B}\ }\textbf {\bibinfo {volume} {83}},\ \bibinfo {pages} {045417}
  (\bibinfo {year} {2011})}\BibitemShut {NoStop}%
\end{thebibliography}%

\end{document}